\documentclass[10pt, journal,transmag]{IEEEtran}

\usepackage{amsmath}
\usepackage{graphicx}
\usepackage{dcolumn}
\usepackage{bm}
\usepackage[utf8]{inputenc}
\usepackage[T1]{fontenc}
\usepackage{mathptmx}
\usepackage{setspace}

\begin{document}

\title{Economic Hysteresis and Its Mathematical Modeling}

\author{{\bf Isaak D. Mayergoyz}\\ECE Department, University of Maryland \\College Park, MD 20742, USA\\ imayergoyz@gmail.com \\
\\
{\bf Can E. Korman} \\ Department of ECE, George Washington University \\ Washington DC 20052, USA \\korman@gwu.edu}

\maketitle

\date{\today}

\begin{abstract}
Hysteresis is treated as a history dependent branching, and the use of the classical Preisach model for the analysis of macroeconomic hysteresis is first discussed. Then, a new Preisach-type model is introduced as a macroeconomic aggregation of more realistic microeconomic hysteresis than in the case of the classical Preisach model. It is demonstrated that this model is endowed with a more general mechanism of branching and may account for the continuous evolution of the economy and its effect on hysteresis. Furthermore, it is shown that the sluggishness of economic recovery is an intrinsic manifestation of hysteresis branching.

\end{abstract}

\renewcommand{\thefootnote}{\alph{footnote}}

\section{\label{sec:level1}Introduction}

It is known that economic activities (such as employment and trade) exhibit hysteresis. This economic hysteresis has a long and instructive history. Over the years many facts related to economic hysteresis were discussed in the literature.\footnote{See, for instance, \cite{blanchard1986}, \cite{fatas2016}, \cite{fatasOct2016}, \cite{blanchard2018}, \cite{cross2014}, \cite{belke1999}, \cite{belke2008}, \cite{crossbook}, \cite{cross1993}, \cite{amable1995}, \cite{piscitelli1999}, \cite{gocke2019}, \cite{gocke2014}, \cite{gocke2002}, \cite{amable1994}, \cite{amable1991}, \cite{baldwin1988}, \cite{baldwin-2911-1989} and \cite{baldwin-2828-1989}} This list of references is not exhaustive but rather suggestive. As of today, a concept of economic hysteresis is still lacking an unambiguous and rigorous definition accepted by the majority of researchers in economics.

During the 1980's, it was perceived that the natural rate hypothesis was inconsistent with observed long-lasting high unemployment rates. As a way out of this predicament, Blanchard and Summers suggested that unemployment rates are history-dependent and exhibit hysteresis \cite{blanchard1986}. They proposed to use the unit root test as a validation of the presence of hysteresis. This paper was very influential in bringing the notion of hysteresis into the center of economic research. However, outstanding questions still remain concerning the origin of economic hysteresis and why this hysteresis is responsible for persistent high unemployment rates.

Another approach to the mathematical modeling of economic hysteresis was initiated during the last 30 years.\footnote{See \cite{cross2014}, \cite{belke1999}, \cite{belke2008}, \cite{crossbook}, \cite{cross1993}, \cite{amable1995}, \cite{piscitelli1999}, \cite{gocke2019}, \cite{gocke2014}, \cite{gocke2002}, \cite{amable1994}, \cite{amable1991}, \cite{baldwin1988}, \cite{baldwin-2911-1989} and \cite{baldwin-2828-1989}} This became possible due to the development of sophisticated mathematical models of hysteresis used in magnetics as well as in other areas of science and technology. The most prominent hysteresis model is the classical Preisach model whose generality was originally stressed in \cite{krasno_book},  \cite{IDMBook1991} and \cite{IDMBook2003}. This model attracted the attention of economists who have started to use it in the study of economic hysteresis. It must be remarked that this economic hysteresis research has been by and large confined to the classical Preisach model of hysteresis.

The purpose of this paper is three fold. First, an attempt is made to provide a clear definition of hysteresis as {\bf history-dependent branching} and to discuss the relevance of this definition to macroeconomic hysteresis. Subsequently, attempts are made to elaborate on the origin of macroeconomic hysteresis. Second, the classical Preisach model of hysteresis is introduced. It is pointed out that its structure is intimately related to the origin of economic hysteresis. Third, a far reaching generalization of the classical Preisach model for macroeconomic hysteresis is presented. This model is constructed as a macroeconomic aggregation of a more realistic microeconomic hysteresis than in the case of the classical Preisach model. In contrast to the classical Preisach model, this model is endowed with a more general and sophisticated mechanism of hysteresis branching. Furthermore, a unique feature of hysteresis in economics in comparison with hysteresis in the natural sciences is due to its ever changing background as a result of continuous economic evolution. It turns out that the generalized Preisach model may be very helpful to account for the continuous evolution of the economy and its effect on hysteresis. Finally, it is also demonstrated within the framework of the Preisach hysteresis modeling that the sluggishness of economic recovery is an intrinsic manifestation of hysteresis branching, and this sluggishness may be predicted by using relevant microeconomic data.

The paper is written to be self-contained and accessible to a broad audience. For this reason, in our discussion of hysteresis models many mathematical details and proofs are omitted. The main emphasis in our exposition is on the conceptual foundations of these models and their economic relevance.

\section{\label{sec:whatishysteresis}What is Hysteresis?}

\begin{figure}[t]
\centerline{\includegraphics[width=6cm]{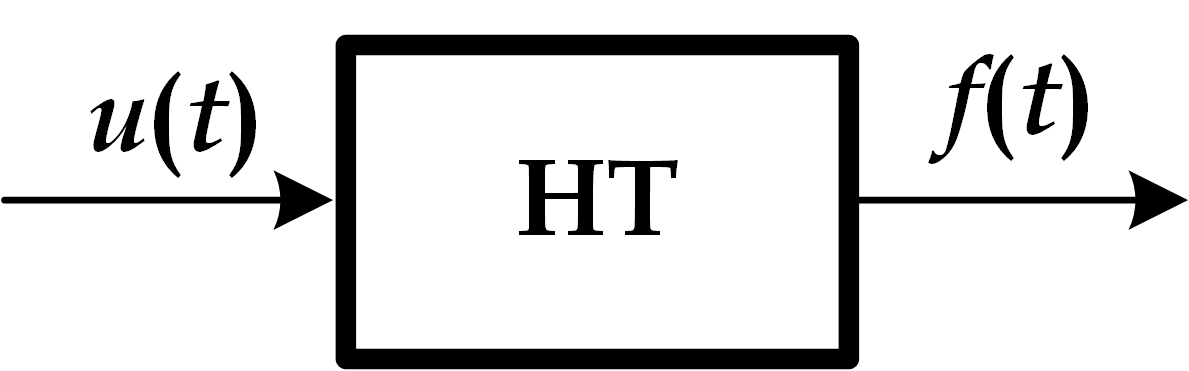}}
\caption{Hysteresis Transducer.}
\label{fig_1}
\end{figure}

Hysteresis phenomenon is ubiquitous. It is encountered in many different areas of science, technology and daily life. Prominent examples of hysteresis include magnetic hysteresis, mechanical hysteresis, superconducting hysteresis, optical hysteresis, electron-beam hysteresis, adsorption hysteresis, economic hysteresis, etc. The phenomenon of hysteresis is also encountered in biology and neuroscience, see for instance \cite{neural2020}. For this reason, it is desirable to study hysteresis in terms of input-output relationships, without specifying their underlying nature. This study should be applicable to any specific manifestation of hysteresis. In order to achieve this goal, we shall adopt the language of system theory. Specifically, we consider a hysteresis transducer (HT) that converts an input $u(t)$ into an output $f(t)$ (see Fig. \ref{fig_1}). For instance, in economics $f(t)$ may be the unemployment rate, investment, trade, etc., whereas, $u(t)$ may be the interest rate, assessed risk, exchange rate, price, etc.

\begin{figure}[t]
\centerline{\includegraphics[width=7cm]{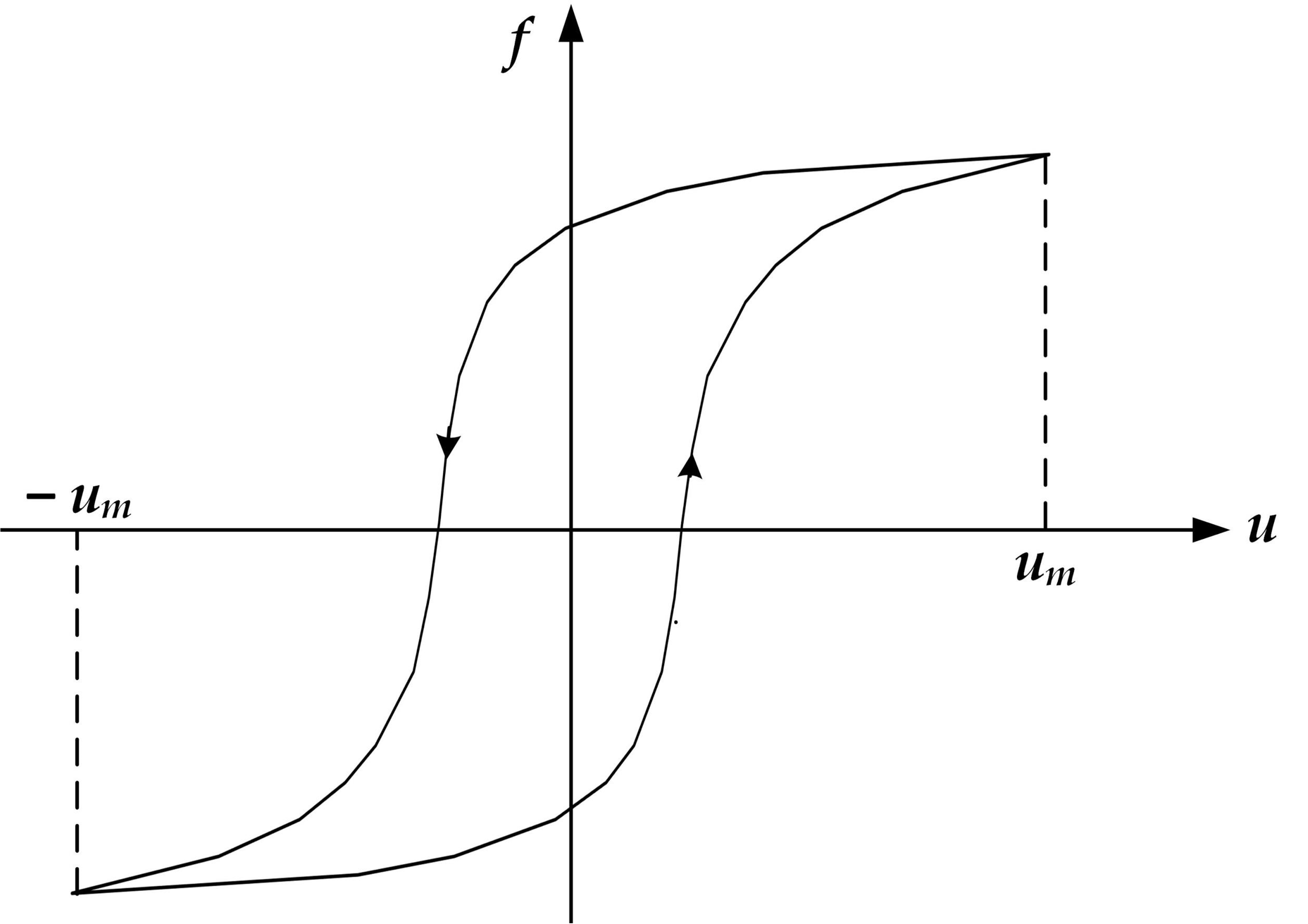}}
\caption{Hysteresis loop.}
\label{fig_2}
\end{figure}

Arguably, the most known and simplest manifestation of hysteresis is a hysteresis loop (see Fig. \ref{fig_2}). Such a hysteresis loop is formed for cyclical variations of input between two extremum values. In particular, for symmetric loops, these extremum values are minimum value $-u_m$ and maximum value $u_m$. Such loops have two branches: an ascending branch corresponding to monotonic increase of input from $-u_m$ to $u_m$ and a descending branch corresponding to monotonic decrease of input from $u_m$ to $-u_m$.

The term “hysteresis” was first introduced by the physicist J. A. Ewing \cite{ewing1881}. This word is of Greek origin with the meaning of “lagging behind”. This is illustrated in Fig. \ref{fig_2}. Indeed, when a symmetric hysteresis loop is traced, variations of output $f(t)$ lag behind the variations of input $u(t)$. For instance, when the descending branch is traced, input $u(t)$ reaches zero, while the output $f(t)$ is still positive. Furthermore, output $f(t)$ reaches zero, whereas input $u(t)$ is already negative. Similar lagging behind variations of output $f(t)$ are observed in tracing the ascending branch of a symmetrical hysteresis loop.

\begin{figure}[t]
\centerline{\includegraphics[width=9.0cm]{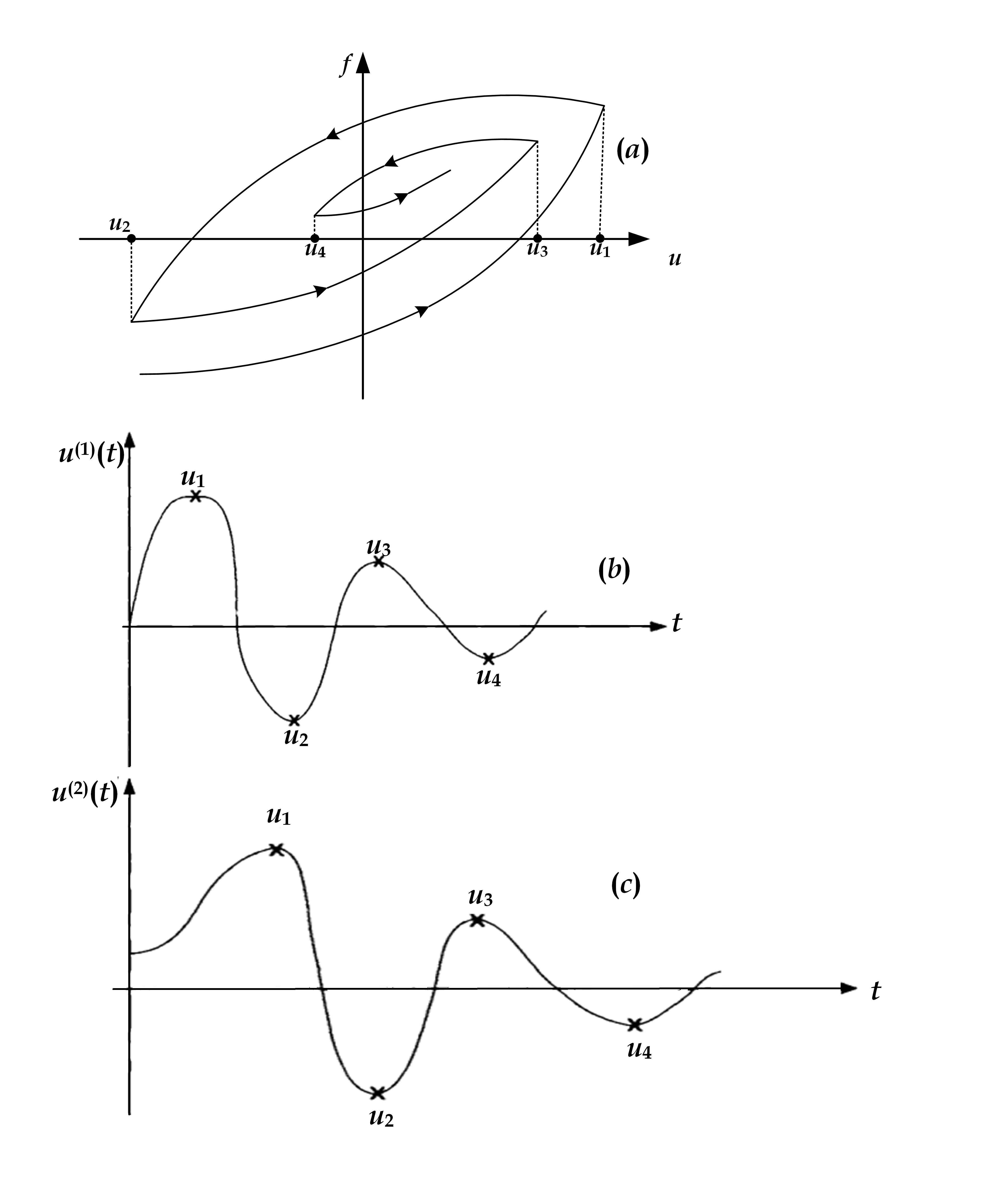}}
\caption{History-dependent multibranch nonlinearity.}
\label{fig_3}
\end{figure}

It is important to point out that hysteresis loops are the simplest manifestations of hysteresis. In fact, hysteresis phenomenon is much more complex and interesting. Its essence can best be described as {\bf history-dependent branching}. In other words, hysteresis can be defined as a multibranch input-output relation for which transitions from one branch to another occur after each input extremum. An example of such a multibranch nonlinearity is shown in Fig. \ref{fig_3}a. It is apparent that the formation of hysteresis loops is a particular case of history-dependent branching. Such a case is realized for cyclical variations of inputs, while branching occurs for arbitrary input variations. To summarize, hysteresis can be understood as an input-output nonlinearity with memory of the input history which reveals itself through branching.

In this paper, we will confine our discussion to {\bf rate-independent hysteresis} where past input memories responsible for branching have relatively simple structures. Namely, the term “rate-independent” means that branches of hysteresis are controlled only by the past extremum values of input, while the speed and particular manner of monotonic input variations between input extremum points has no effect on branching. This is illustrated by Figs. \ref{fig_3}a, \ref{fig_3}b and \ref{fig_3}c. Here, Figs. \ref{fig_3}b and \ref{fig_3}c represent two different inputs $u^{(1)} (t)$ and $u^{(2)} (t)$ that successively assume the same extremum values $u_1$, $u_2$, $u_3$ and $u_4$ but vary in time differently between these values. Then, for rate-independent hysteresis these two inputs will result in the same branching (see Fig. \ref{fig_3}a) provided that the initial state of the system is the same. This implies that rate-independent hysteresis is endowed with a discrete memory structure which consists of past-extremum values of input.

\begin{figure}[t]
\centerline{\includegraphics[width=7cm]{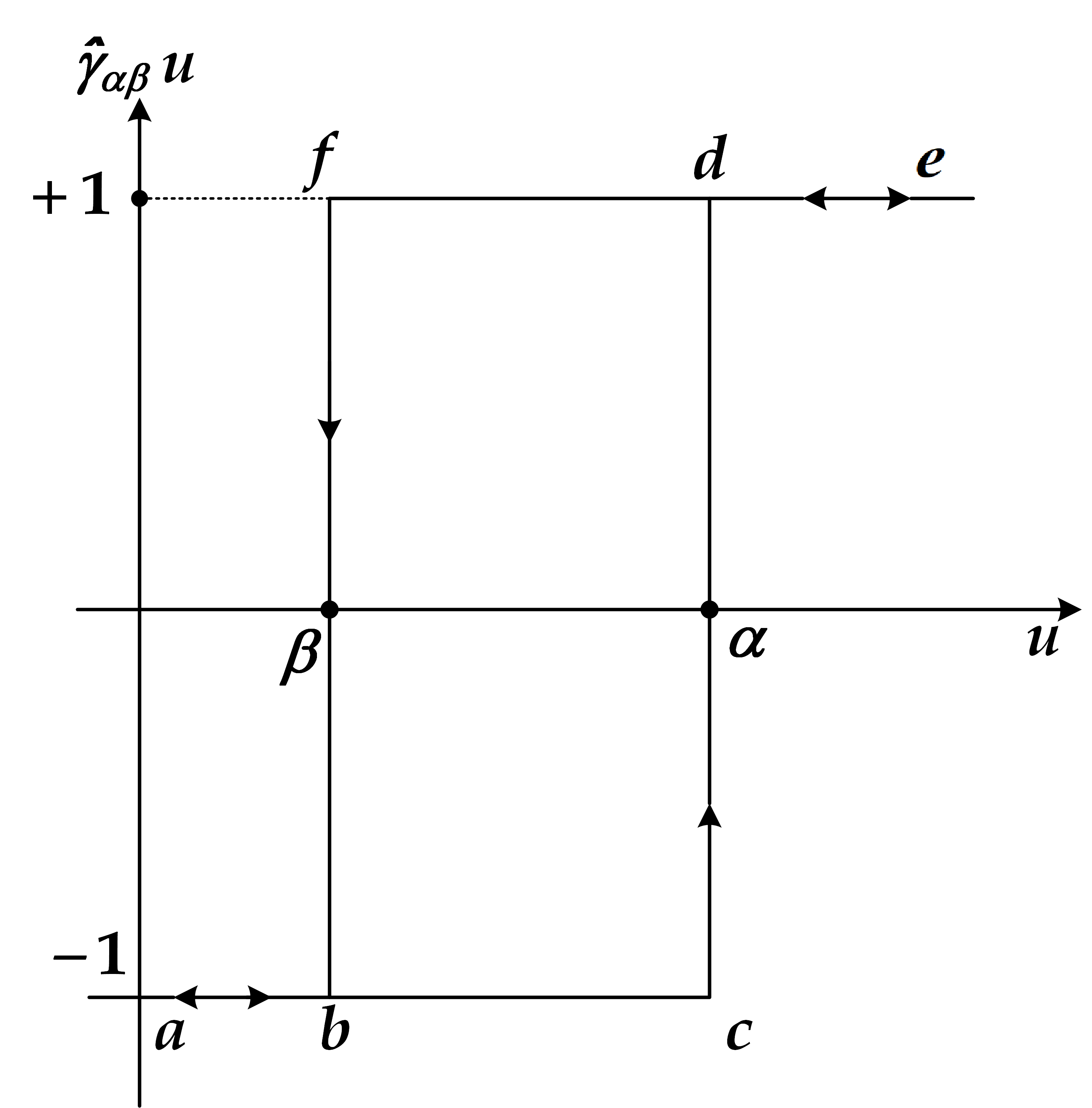}}
\caption{Basic rectangular hysteresis loop.}
\label{fig_4}
\end{figure}

The given definition of hysteresis as history-dependent branching is consistent with observed economic activities which typically consist of numerous upward and downward output variations. These output variations as a function of input can be viewed as branches that sequentially follow one another in time. This sequential branching can be graphically illustrated by Fig. \ref{fig_3}a. Furthermore, it is important to point out that the notion of rate-independent hysteresis is naturally applicable in economics. This is because of high economic inertia related to the huge scale of the economy as well as the innate human nature to maintain stability by resisting fast changes.

The notion of hysteresis as a multibranch nonlinearity was introduced in \cite{IDMPhysRev1986} and \cite{IDMTransMag1986}. This notion has been gradually accepted in the economic literature. For instance, it is stated in \cite{gocke2002} that "{\it all types of 'genuine' hysteresis show the common feature of multibranch non-linearity.}"

It is clear from the previous discussion that the main challenge in the development of rigorous hysteresis models is the creation of special mathematical tools. These tools are expected to detect and accumulate past extremum values of input and then choose branches of hysteresis nonlinearity according to the stored history of past input extremum values. Such hysteresis models should be endowed with the {\it predictive} power of branching which should be the main purpose of hysteresis modeling in the first place, and can be potentially useful for economic predictions.

Another fundamental question in the study of hysteresis is the nature of its origin. The origin of hysteresis in physical systems is very often related to the multiplicity of metastable states in such systems. These metastable states correspond to various local energy minima. In this sense, the origin of hysteresis in physical systems is somewhat remote from the mathematical structure of Preisach models of hysteresis discussed below. In contrast, the origin of economic hysteresis is of a different nature. This origin is due to the occurrence of numerous microeconomic binary actions, such as hiring-firing in the case of hysteresis of cyclical unemployment, or buying-selling in the case of trade or stock market hysteresis. Each of these binary actions may be triggered at different values of economic input, such as price, interest rates, assessed risk in the market, etc. Remarkably, the aforementioned origin of economic hysteresis is intimately related to and reflected in the mathematical structure of Preisach models, in which the microeconomic binary actions are represented by rectangular loops (see Fig. \ref{fig_4}). These models of hysteresis are subsequently discussed in this paper.

\section{\label{sec:level1}Classical Preisach Model}

The inception of the classical Preisach model can be traced back to the paper of Preisach "On the Magnetic Aftereffect" \cite{preisach1935}. The English translation of this paper has also recently appeared in \cite{preisach2017}. It is worthwhile to mention that the classical paper of Preisach is mostly concerned with aftereffect phenomena, while the discussion of hysteresis modeling is very brief. Since this paper dealt exclusively with magnetic hysteresis, the Preisach model was first regarded as a physical model, and it was the focus of considerable research in the field of magnetics for many years.\footnote{See, for instance, \cite{neel}, \cite{biorci425}, \cite{brown} and \cite{damla}} Somewhat later, the Preisach model was independently discovered and studied for adsorption hysteresis by Everett \cite{everett749}. This work clearly demonstrated the generality of the Preisach model, and showed that its applicability was not limited to specific areas of physical phenomena.

The next significant step in the development of the Preisach model occurred in the late 1970s and the early 1980s. In the works of M. Krasnoselskii, A. Pokrovskii and I. Mayergoyz it was shown that the Preisach model contained a new mathematical idea and can be stated without any reference to underlining physical phenomena. As a result, the Preisach model was separated from its physical connotation and represented in purely mathematical terms that are similar to the spectral decomposition of operators as described in \cite{friedman}. Accordingly, a new mathematical tool has evolved that can be employed for the mathematical description of hysteresis of various nature, physical or otherwise, including hysteresis in economics.

The general mathematical definition of the Preisach model can be outlined as follows. Consider an infinite set of simplest hysteresis operators $\hat{\gamma}_{\alpha\beta}$. Each such operator can be represented by a rectangular loop on the input-output diagram (see Fig. \ref{fig_4}). This counter-clockwise loop is characterized by the numbers $\alpha$ and $\beta$ which correspond to “up” and “down” switching (triggering) values of input, respectively. Here, we note that $\alpha\geq\beta$. For instance, this counter-clockwise hysteresis loop may be used to describe unemployment hysteresis. Namely, the switching up value $\alpha$ corresponds to the triggering threshold for firing and unemployment increase, while the switching down value $\beta$ corresponds to the triggering threshold for hiring and unemployment decrease. The difference $(\alpha -\beta )$ can be viewed as a range-of-inactivity which is also called "band-of-inactivity" or "hysteresis-band" in \cite{baldwin1988} and \cite{baldwin-2911-1989}.

There are also many cases in economics and other fields where a clockwise loop (with $\alpha\leq\beta$) may be useful to employ (see, for instance \cite{gocke2019}). However, in the subsequent discussions and without any loss of generality, we will only consider counter-clockwise loops. Outputs of these elementary hysteresis operators may assume only two values, $+1$ and $-1$, although other binary combinations of outputs (for instance, $1$ and $0$) may just as easily be considered as well.  In other words, these operators can be interpreted as two-position switches with “up” and “down” positions, respectively:
\begin{equation}
\hat{\gamma}_{\alpha\beta} u(t) = +1 \ ,
\label{eq1.1}
\end{equation}
\noindent
and
\begin{equation}
\hat{\gamma}_{\alpha\beta} u(t) = -1 \ .
\label{eq1.2}
\end{equation}

\noindent
As the input, $u(t)$, is monotonically increased, the ascending branch $abcde$ is followed. When the input is monotonically decreased, the descending branch $edfba$ is traced. It is worthwhile noting that the operators $\hat{\gamma}_{\alpha\beta}$ represent the simplest hysteresis nonlinearities with {\bf local (Markovian) memories}. The latter implies that the output value at any instant of time completely accounts for the past history.

Along with the set of rectangular loops $\hat{\gamma}_{\alpha\beta}$, we consider a weight function $\mu (\alpha , \beta )$ that is generally referred to as the Preisach function. Now, the Preisach model can be defined as follows:
\begin{equation}
f(t) =
\iint\limits_{\alpha\geq\beta}
\mu (\alpha, \beta)
\hat{\gamma}_{\alpha\beta} u(t) d\alpha d\beta
.
\label{eq1.3}
\end{equation}

\noindent

\begin{figure}[t]
\centerline{\includegraphics[width=7cm]{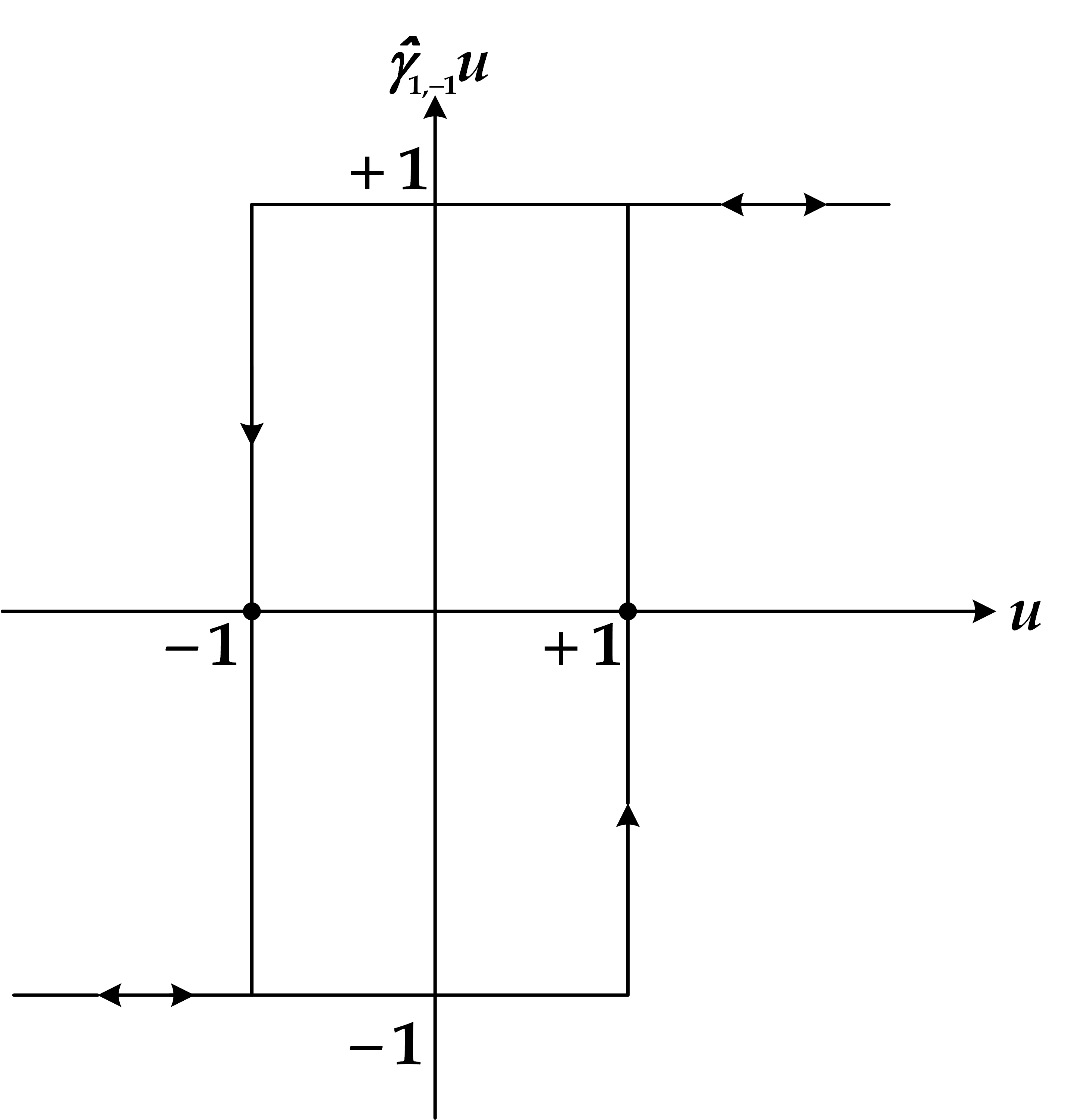}}
\caption{Rectangular "mother" hysteresis loop.}
\label{fig_5}
\end{figure}

From the aforementioned description, it is clear that the Preisach model (\ref{eq1.3}) is constructed as a superposition of the simplest rectangular hysteresis loops $\hat{\gamma}_{\alpha, \beta}$. These rectangular loops can be interpreted as the primary building blocks for the Preisach model (\ref{eq1.3}). The notion that a complex operator can be represented as a superposition of simplest operators is not entirely new, and this idea was exploited before in mathematics, particularly in functional analysis \cite{friedman}, \cite{neumann}. For example, according to the spectral decomposition theory for self-adjoint (Hermitian) operators, any self-adjoint operator can be represented as a superposition of projection operators that are, in a way, the simplest self-adjoint operators. This analogy reveals that from the mathematical point of view the Preisach model (\ref{eq1.3}) can be interpreted as a spectral decomposition of complex hysteresis operators into the simplest rectangular hysteresis loop operators $\hat{\gamma}_{\alpha, \beta}$. Furthermore, there is also a noteworthy parallel between the Preisach model and wavelet transforms which are commonly used in the field of signal processing. In fact, all rectangular loop operators $\hat{\gamma}_{\alpha, \beta}$ can be obtained by translating and dilating the rectangular loop operator, $\hat{\gamma}_{1, -1}$, (see Fig. \ref{fig_5}), that can be regarded as the “mother loop operator.” In this sense, the Preisach model can be viewed as a “wavelet operator transform.”

From the above discussion it is clear that the Preisach model has been defined without any reference to a particular origin of hysteresis. This clearly reveals the phenomenological nature of the model and its mathematical generality. This suggests that it can be used for the description of hysteresis of any nature, including hysteresis in economic systems. Indeed, economists have applied the classical Preisach model to the analysis of economic systems since the early 1990s (see the works of \cite{belke1999}, \cite{cross1993} and \cite{amable1995}).

As discussed earlier, in economics rectangular loops can be used as mathematical representations of {\it binary} microeconomic activities, such as buying-selling  or hiring-firing decisions. In this sense, the Preisach model may be construed as an aggregation of numerous binary microeconomic actions with various switching (triggering) values driving economic activities in the market. Therefore, it is clear that the mathematical structure of the Preisach model is directly related to the origin of macroeconomic hysteresis. Indeed, the basic rectangular loops have some clear relevance to basic microeconomic activity and decision making. This adds further credence to the Preisach approach of modeling macroeconomic activities where piece-wise monotonic branching occurs in the market due to the binary actions of a very large number of agents all driven by a collective risk with different individual risk tolerances.

In economics literature, the basic rectangular loops are viewed as microeconomic hysteresis of individual firms involved in trade or employment. Such hysteresis is often referred to as "weak" hysteresis (see \cite{amable1991}). On the other hand, the hysteresis described by the classical Preisach model is viewed as an aggregation of microeconomic hysteresis to the macroeconomic level. This macroeconomic hysteresis is often referred to as "strong" hysteresis (see \cite{amable1991}).

It is interesting to point out that the justification for the representation of microeconomic hysteresis by rectangular loops can be done by using the Random Field Ising Model (RFIM). This model is employed to account for the interaction between individual firms (or economic agents). This approach is discussed in the very interesting paper of Bouchaud \cite{bouchaud2013}. A similar discussion on the origin of rectangular hysteresis loops in magnetics in connection with Barkhausen jumps is presented in the publications of Sethna and co-workers \cite{sethna1993}, \cite{dahmen1996} and \cite{sethna-book}.

It is apparent from formula (\ref{eq1.3}) that the structure of the Preisach model has two components: rectangular loops $\hat{\gamma}_{\alpha, \beta}$ and weight function $\mu (\alpha, \beta)$. In magnetics and other areas of science and engineering, the identification of the weight function $\mu (\alpha ,\beta )$ is performed by using the so-called first order reversal curves (FORC's). These curves are found by performing {\it controlled} experiments for wide ranges of input variations. Such experiments are not realistically possible in economic environments. This brings the question of the meaning of the weight function in the context of economics. It turns out that in the case of unemployment hysteresis $\mu (\alpha ,\beta )$ is directly related to the overall employment capacity of binary agents (firms) with the same triggering values of $\alpha$ and $\beta$. Indeed, due to the finite number of binary agents, the following formula is valid:
\begin{equation}
\mu (\alpha, \beta)  =
\sum\limits_{k}
\nu (\alpha, \beta)
\delta (\alpha -\alpha_k ) \delta (\beta -\beta_k )
,
\label{sum_mu}
\end{equation}
where $\delta$ is the traditional notation for the Dirac delta function.

After substituting the last expression into formula (\ref{eq1.3}), we arrive at
\begin{equation}
f(t) =
\sum\limits_{k}
\nu (\alpha_k , \beta_k )
\hat{\gamma}_{\alpha_k \beta_k } u(t)
.
\label{sum_output}
\end{equation}
Now, it is clear that $\nu (\alpha_k , \beta_k )$ can be viewed as the overall employment capacity of binary agents with the same triggering values $\alpha_k$ and $\beta_k$. The information on this capacity requires the collection of proper data on binary agents and their employment activities.

In our discussion of microeconomic hysteresis, the binary actions of hiring-firing have been idealized by assuming that they occur {\it abruptly} (see Fig. \ref{fig_4}). In reality, these actions may occur gradually. {\bf A more realistic representation of microeconomic hysteresis, which leads to the generalized Preisach model, is discussed in Section \ref{sec:generalized}}.

It is also clear from formula (\ref{eq1.3}), as well as formula (\ref{sum_output}), that the structure of macroeconomic hysteresis is determined by the weight function $\mu (\alpha, \beta)$. Since in the classical Preisach model this weight function depends only on $\alpha$ and $\beta$, this implies that the structure of macroeconomic hysteresis is {\it fixed} regardless of the variation of input $u(t)$. However, modern economies continuously evolve, sometimes resulting in fundamental changes in their underlying structures. This suggests that the structure of economic hysteresis evolves with time, too. Such economic evolutions should be taken into account by exploring the possible dependence of the weight function on input variations. This can be accomplished by considering the generalized Preisach model discussed in  Section \ref{sec:generalized}.

\section{\label{sec:level1}Properties of the Classical Preisach Model}

\begin{figure}[t]
\centerline{\includegraphics[width=7cm]{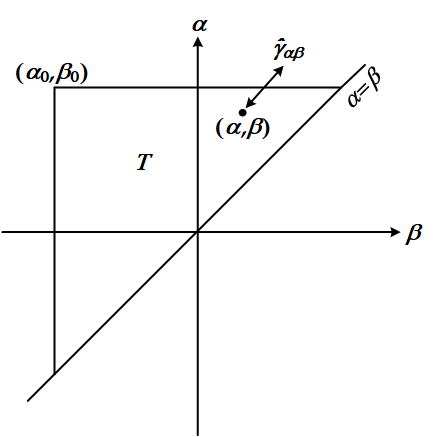}}
\caption{Geometric interpretation of the Preisach model.}
\label{fig_6}
\end{figure}

Having defined the Preisach model by formula (\ref{eq1.3}), the question of how the Preisach model works must be addressed. In other words, it needs to be understood how the Preisach model (\ref{eq1.3}) detects past extremum values of input, stores them in its structure and chooses the appropriate branches of hysteresis nonlinearity according to accumulated histories. The above question can be fully answered by using a special diagram technique which is based on the following geometric interpretation of the Preisach model. This geometric interpretation is based on the observation that there is a one-to-one correspondence between rectangular loop operators $\hat{\gamma}_{\alpha, \beta}$ and points $(\alpha , \beta )$ of the half-plane $\alpha\geq \beta$ (see Fig. \ref{fig_6}). In other words, each point of the half-plane $\alpha\geq \beta$ can be identified with only one particular rectangular loop whose “up” and “down” switching values are respectively equal to $\alpha$ and $\beta$ coordinates of the point.

\begin{figure}[t]
\centerline{\includegraphics[width=7cm]{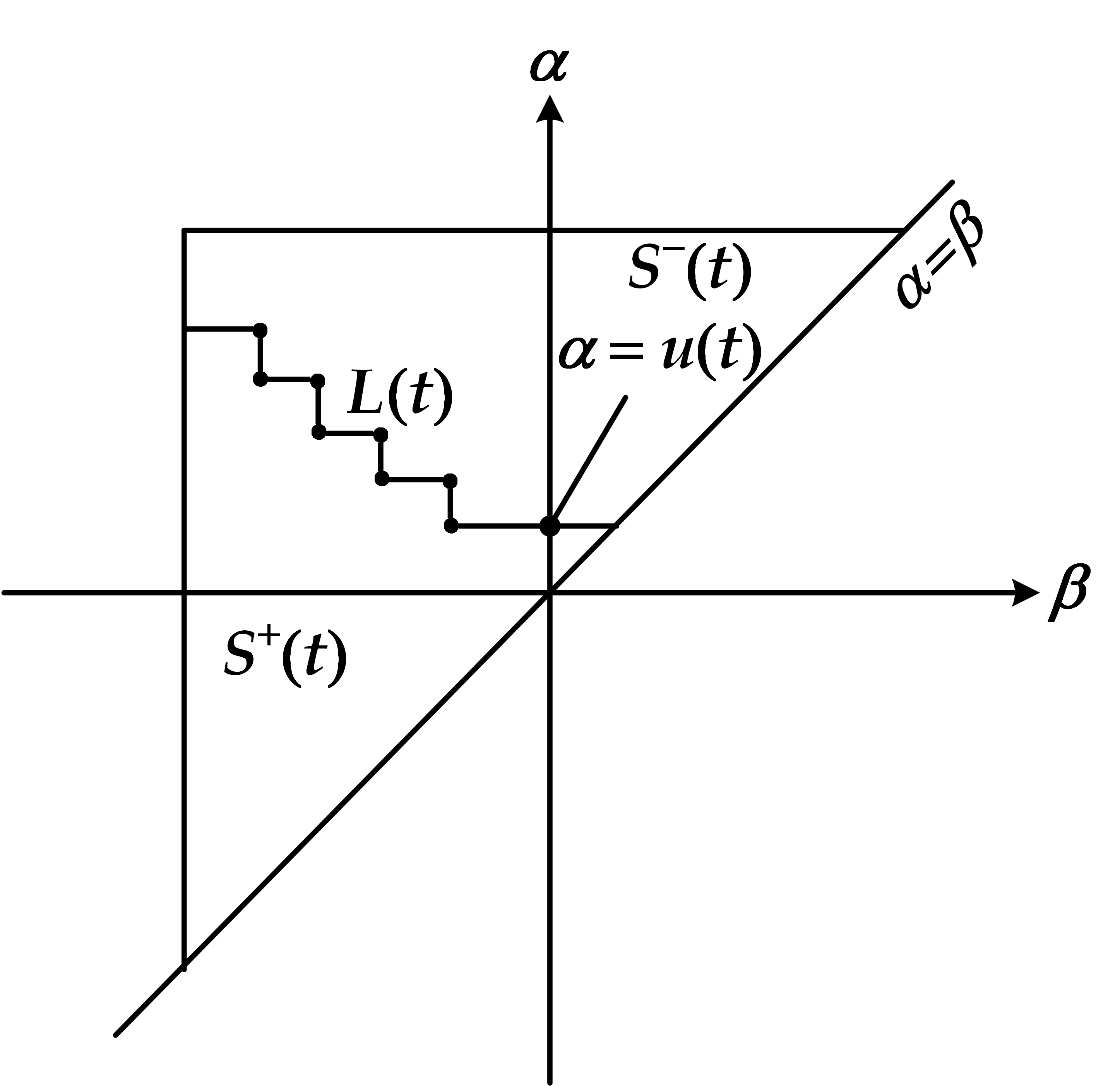}}
\caption{Staircase interface $L(t)$ for increasing input.}
\label{fig_7}
\end{figure}

\begin{figure}[t]
\centerline{\includegraphics[width=7cm]{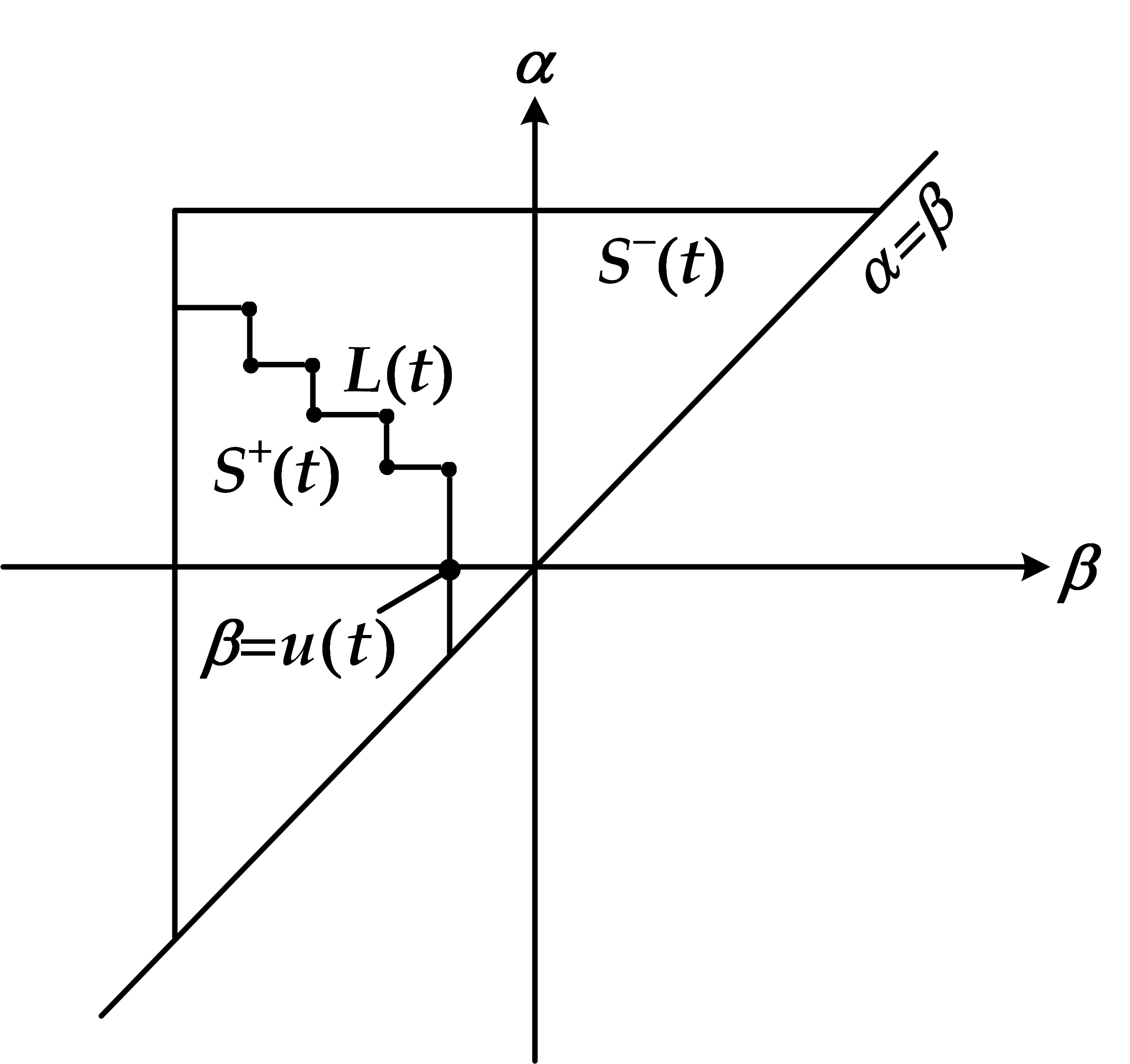}}
\caption{Staircase interface $L(t)$ for decreasing input.}
\label{fig_8}
\end{figure}

Let us assume that the weight function is non-zero within some triangle $T$ (see Fig. \ref{fig_6}). (In the case of economic hysteresis when $u(t)$ is always positive, triangle $T$ is always within the first quadrant.) Now, it can be shown that at any instant of time, the triangle $T$ is subdivided into two sets: $S^{+} (t)$ consisting of points $(\alpha, \beta)$ for which the corresponding rectangular loops $\hat{\gamma}_{\alpha\beta}$ are in the “up” position, and $S^{-} (t)$ consisting of points for which the corresponding rectangular loops $\hat{\gamma}_{\alpha\beta}$ are in the “down” position. The structures of $S^{+} (t)$ and $S^{-} (t)$ can be determined by the following two facts: (i) if the input is monotonically increased, then a horizontal link is formed at the interface between $S^{+} (t)$ and $S^{-} (t)$, and this link moves upwards until the input reaches its maximum value, (ii) on the other hand, if the monotonic input increase is followed by a monotonic decrease, then a vertical link of the interface is formed, and this link moves from right to left until the input reaches its minimum value. The validity of these two facts is directly based on the definition of the rectangular loop operators $\hat{\gamma}_{\alpha\beta}$ and their geometric interpretation. The above two facts imply that the interface between the positive and negative sets is a staircase line, $L(t)$, consisting of horizontal and vertical links. The vertices of this staircase line have $(\alpha ,\beta )$ coordinates that are the past maximum and minimum values of input, respectively (see Figs. \ref{fig_7} and \ref{fig_8}). It is also apparent that the final link of $L(t)$ is attached to the line $\alpha = \beta$ and it moves when the input is changed. This final link is a horizontal one and it moves upwards as the input is increased (see Fig. \ref{fig_7}), whereas the final link is a vertical one and it moves from right to left as the input is decreased (see Fig. \ref{fig_8}).

Therefore, at any instant of time the integral in formula (\ref{eq1.3}) can be subdivided into two integrals over $S^{+} (t)$ and $S^{-} (t)$, respectively. Since,
\begin{equation}
\hat{\gamma}_{\alpha, \beta} u(t) = +1,
\ \ {\rm if}\ (\alpha, \beta) \in {S^{+} (t)}
\label{eq1.5}
\end{equation}
and
\begin{equation}
\hat{\gamma}_{\alpha, \beta} u(t) = -1,
\ \ {\rm if}\ (\alpha, \beta) \in {S^{-} (t)}
,
\label{eq1.6}
\end{equation}
from formula (\ref{eq1.3}) we find
\begin{equation}
f(t) =
\iint\limits_{S^{+} (t)}
\mu (\alpha, \beta)
d\alpha d\beta
-
\iint\limits_{S^{-} (t)}
\mu (\alpha, \beta)
d\alpha d\beta
.
\label{eq1.7}
\end{equation}
\noindent
From expression (\ref{eq1.7}), it can be seen that the instantaneous value of output depends on a particular subdivision of the triangle $T$ into positive and negative sets $S^{+} (t)$ and $S^{-} (t)$. This subdivision is determined by the particular shape of the interface $L(t)$. In turn, this shape depends on the past extremum values of input because these extremum values define coordinates of the vertices of $L(t)$. As a result, the past extremum values of input shape the staircase interface $L(t)$ and, accordingly, leave their mark upon future values of output $f(t)$ and, consequently, future hysteresis branching.

This branching occurs because a monotonic input increase results in the formation of a horizontal final link of $L(t)$ and its upward movement, while a monotonic input decrease results in the formation of a vertical final link of $L(t)$ and its leftward movement. {\bf These two different modifications of} $L(t)$ {\bf result in the ascending and descending output branches for monotonically increasing and decreasing inputs, respectively. This type of branching is the essence of hysteresis}.

The presented discussion clearly reveals that a new qualitative property of {\bf non-local (non-Markovian) memory} consisting of past input extrema emerges as a result of the superposition of rectangular loops with {\bf local (Markovian) memories}. In economic terms, this means that the new qualitative property of non-local memory emerges at the macroeconomic level due to the aggregation of microeconomic rectangular hysteresis loops with different ranges of inactions. The emergence of the above qualitative property of nonlocal memory justifies the notion of macroeconomic hysteresis as "strong" hysteresis.

It turns out that the Preisach model does not accumulate all past extremum values of input. Some of them can be erased by subsequent input variations. This fact can be stated as follows:

\medskip\noindent
{\bf ERASURE PROPERTY} {\it Each local input maximum erases the vertices of $L(t)$ whose $\alpha$-coordinates are below this maximum, and each local minimum erases the vertices whose $\beta$-coordinates are above this minimum.}

The erasure property is asserted above in purely geometric terms. This makes this property quite transparent. However, the same property can also be described in analytical terms. The analytical formulation complements the geometric one because it is directly phrased in terms of time input variations as follows:

\medskip\noindent
{\bf ERASURE PROPERTY} {\it Only the sequence of alternating dominant input extrema are stored by the Preisach model. All other input extrema are erased.}

\begin{figure}[t]
\centerline{\includegraphics[width=8.5cm]{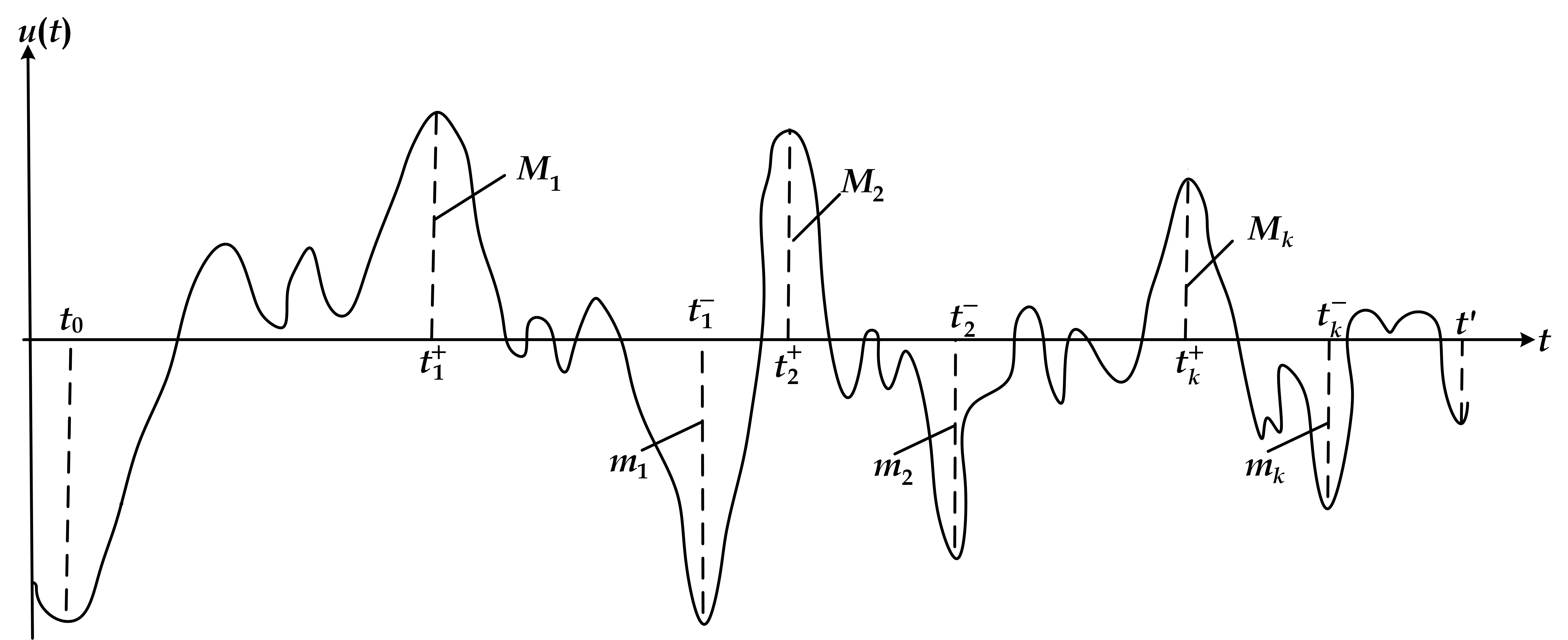}}
\caption{Sequence of dominant input extrema.}
\label{fig_8-9}
\end{figure}

\begin{figure}[t]
\centerline{\includegraphics[width=7cm]{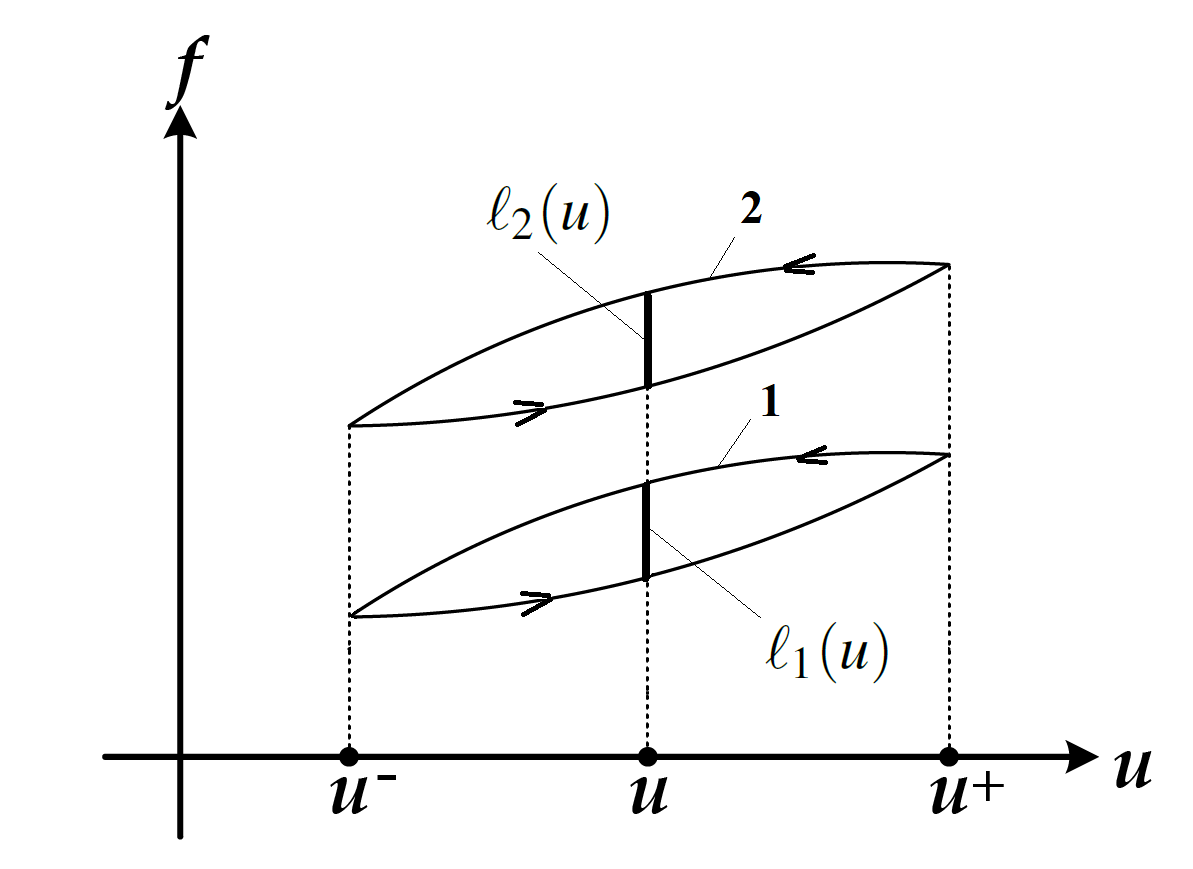}}
\caption{Geometric illustration of the Congruency Property.}
\label{fig_9}
\end{figure}

This statement of the erasure property is illustrated in Fig. \ref{fig_8-9}, where dominant input extrema are marked as $\{ M_1 , m_1 , M_2 , m_2 , \ldots , M_k , m_k , \ldots \}$. Only these dominant input extrema will define the $(\alpha , \beta )$ coordinates of the staircase interface vertices (see Figs. \ref{fig_7} and \ref{fig_8}). All other input extrema will be erased by these dominant input extrema. It is clear from Fig. \ref{fig_8-9} that the first dominant extremum coincides with the global maximum $M_1$, while the second dominant extremum coincides with the global minimum $m_1$ in the time interval $t>t_{1}^{+}$. Similarly, the third dominant extremum $M_2$ is the global maximum in the time interval $t>t_{1}^{-}$, while the fourth dominant extremum $m_2$ is the global minimum in time interval $t>t_{2}^{+}$. Subsequent global extrema are defined in the same way.

Let us next review another characteristic property of the Preisach model that is valid for cyclical input variations, which has direct relevance to economic cycles. Consider $u_1 (t)$ and $u_2 (t)$ to be two inputs that may have different past histories, that is different alternating dominant extrema. However, starting from some instant of time, these inputs vary back and forth between the same two consecutive extremum values, $u^+$ and $u^-$. It can be shown that these periodic input variations result in hysteresis loops that are congruent, as shown in Fig. \ref{fig_9}. This means that the coincidence of these loops can be achieved by the appropriate translation of these loops along the $f$-axis. This property can be stated as follows:

\medskip\noindent
{\bf CONGRUENCY PROPERTY} {\it All  hysteresis loops corresponding to back-and-forth variations of inputs between the same two consecutive extremum values are congruent.}
\medskip

Given the phenomenological nature of the Preisach model and the fact that it can be used independently of the underlying specifics of the system, the question arises concerning the conditions of its applicability. It can be shown that the aforementioned Erasure Property and the Congruency Property constitute the necessary and sufficient conditions for the representation of actual hysteresis nonlinearities by the Preisach model (for the proof see \cite{IDMBook1991} and \cite{IDMPhysRev1986}). This can be stated as follows:

\medskip\noindent
{\bf REPRESENTATION THEOREM} {\it The erasure property and the congruency property constitute the necessary and sufficient conditions for the representation of an actual hysteresis nonlinearity by the Preisach model for piece-wise monotonic inputs.}

In natural sciences and engineering applications, the stated theorem clearly establishes the limits of applicability of the classical Preisach model. Within these limits, the classical Preisach model has predictive powers for history-dependent branching. In the case of economics, when the classical Preisach model is reduced to formula (\ref{sum_output}), the Erasure and Congruency properties are automatically satisfied. Consequently, the classical Preisach model of economic hysteresis may accurately predict the future branching on the basis of accumulated past history of input extremum values.

It turns out that the Congruency Property has very important implications in economics, which do not exist in the case of engineering and natural sciences. To demonstrate these implications, let us consider two unemployment cycles represented by two hysteresis loops shown in Fig. \ref{fig_9}. Here, output $f(t)$ can be viewed as a cyclical unemployment rate, whereas input $u(t)$ can be viewed as the interest rate controllable through monetary policy. Then, the length of vertical chords of hysteresis loops can be interpreted as a measure of sluggishness of economic recovery. This measure of sluggishness is {\bf intrinsic} and is a {\bf direct consequence of hysteresis branching}. Indeed, the descending branch of any hysteresis loop, which represents recovery, is always above the ascending branch, which represents economic contraction (see Fig. \ref{fig_9}). This is because the Preisach weight function is always positive due to its microeconomic interpretation as discussed earlier.

\begin{figure}[t]
\centerline{\includegraphics[width=7.5cm]{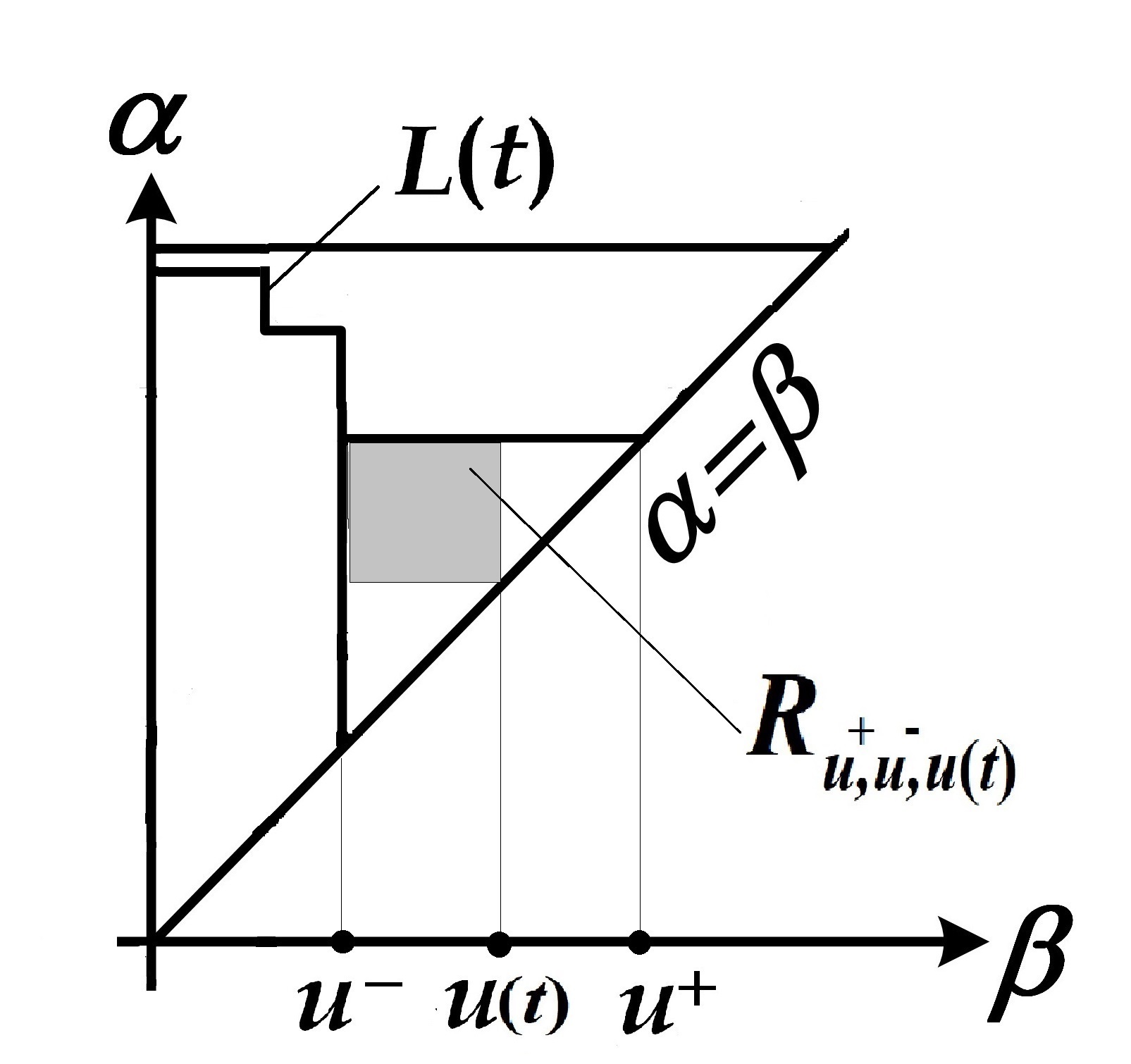}}
\caption{Illustration of rectangle $R_{u^{+},u^{-},u(t)}$ on the Preisach plane.}
\label{fig_rect}
\end{figure}

By using the diagram technique, it is easy to demonstrate that the length of the vertical chord $\ell$ is given by the following formula:
\begin{equation}
\ell =
2 \iint\limits_{R_{u^{+},u^{-},u(t)}}
\mu (\alpha, \beta)
d\alpha d\beta
,
\label{vchord}
\end{equation}
where $R_{u^{+},u^{-},u(t)}$ is the rectangle shown in Fig. \ref{fig_rect}. By employing the meaning of the function $\mu$ in economics, (see formula (\ref{sum_mu})), the last integral can be evaluated in terms of microeconomic data on binary agents whose triggering values $\alpha$ and $\beta$ are such that the points $(\alpha ,\beta )$ are confined within the rectangle $R_{u^{+},u^{-},u(t)}$. Consequently, the prior knowledge of microeconomic employment data may be used to predict the sluggishness of the recovery.

According to the Congruency Property (see Fig. \ref{fig_9}), the vertical chords are the same for all loops corresponding to the same consecutive extremum values $u^{-}$ and $u^{+}$:
\begin{equation}
\ell_1 (u) =
\ell_2 (u)
.
\label{vchord_a}
\end{equation}
This means that the sluggishness of economic recovery does not depend on the history of input variations prior to the commencement of the economic cycle determined by $u^{-}$ and $u^{+}$. In this respect, the sluggishness of economic recovery is {\bf universal}. Remarkably, this universality exists despite the fact that the output values corresponding to $u^{-}$ and $u^{+}$ are history-dependent.

The above discussion clearly reveals the economic importance of hysteresis branching. This branching can be constructed by using the time series traditionally involved in macroeconomic studies. Indeed, by using the time series for output $f(t)$ and input $u(t)$ and by relating $f$ to $u$ at the same time instances, the $f$ vs. $u$ relation can be constructed. This $f$ vs. $u$ relation will exhibit hysteresis branching.

\section{\label{sec:generalized}Generalized Preisach Model}

\begin{figure}[t]
\centerline{\includegraphics[width=7.5cm]{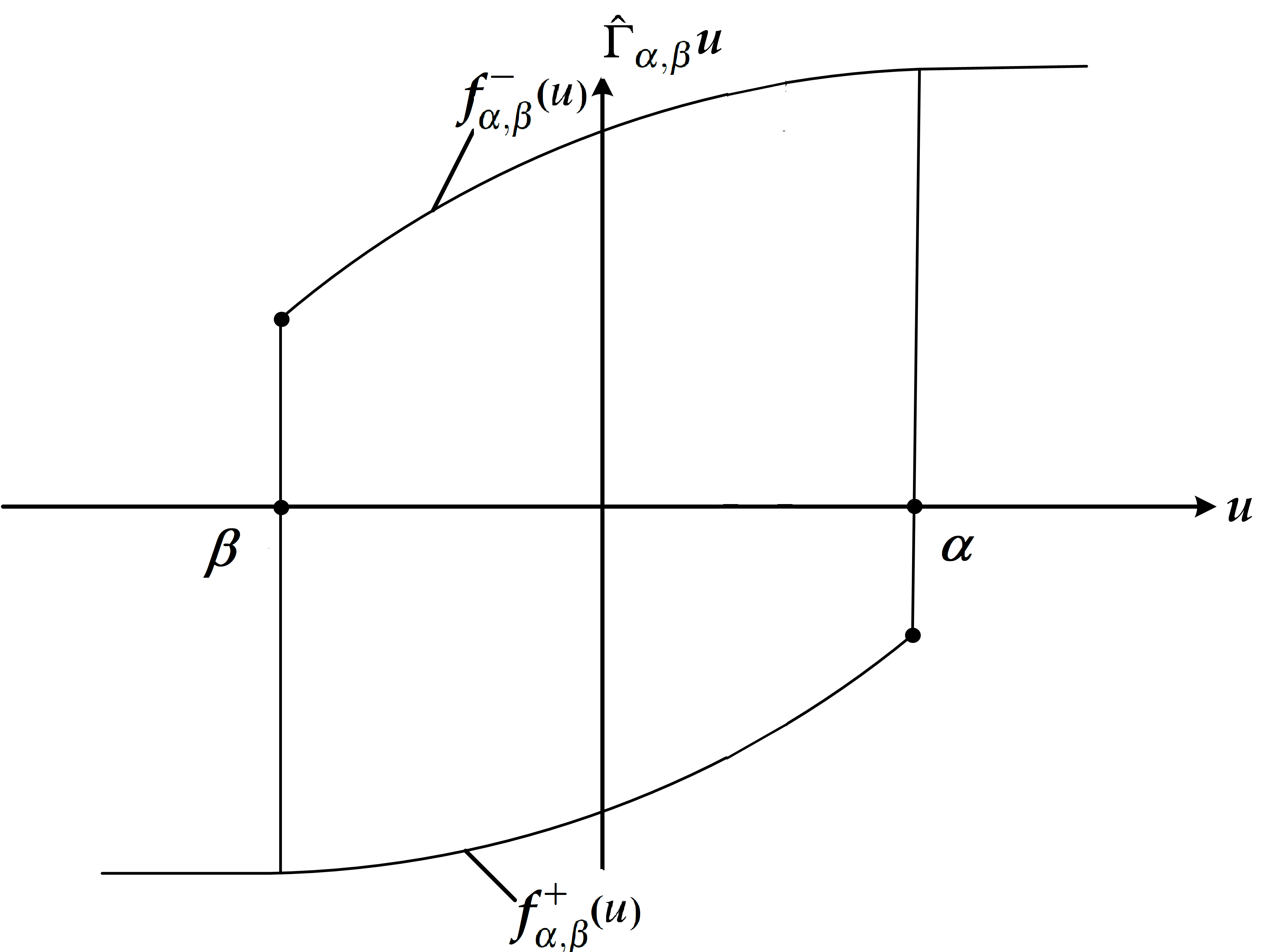}}
\caption{A more realistic binary agent for microeconomic hysteresis.}
\label{fig_stoner}
\end{figure}

A very instructive and informative way to arrive at the generalized Preisach model is through the process of macroeconomic aggregation. According to the classical Preisach model, macroeconomic hysteresis has been construed as the aggregation of rectangular loops representing microeconomic hysteresis. Such loops represent an idealization of microeconomic hysteresis where buying-selling, hiring-firing, etc. are represented in terms of only two possible outputs. A more realistic way to depict microeconomic hysteresis is illustrated by Fig. \ref{fig_stoner}, where $\alpha$ and $\beta$ are  the "up" and "down" triggering values between the ascending branch $f_{\alpha ,\beta}^{+} (u)$ and the descending branch $f_{\alpha ,\beta}^{-} (u)$ of this loop, respectively. It is clear that the hysteresis loop shown in Fig. \ref{fig_stoner} is a generalization of rectangular loops by replacing the range-of-inactivity by the range of gradual activity. The latter may be a more realistic representation of hiring-firing processes. Furthermore, in the case of unemployment hysteresis $f_{\alpha ,\beta}^{+} (u)$ and $f_{\alpha ,\beta}^{-} (u)$ represent the microeconomic employment capacity of individual firms.

The hysteresis loop shown in Fig. \ref{fig_stoner} has a salient feature of representing two reversible parts of microeconomic hysteresis by nonconstant monotonic functions $f_{\alpha ,\beta}^{+} (u)$ and $f_{\alpha ,\beta}^{-} (u)$. This feature may allow one to account for the reversible actions of individual economic agents in a more realistic way. In the case of unemployment hysteresis, for instance, the rectangular hysteresis loops can only represent hiring and firing, which are irreversible around the triggering values. Whereas, the type of hysteresis loops shown in Fig. \ref{fig_stoner} can model a much larger array of reversible employer-employee actions in response to varying conditions. Consequently, through the aggregation process, these types of microeconomic hysteresis loops may lead to a more realistic approach to the modeling macroeconomic hysteresis.

It is interesting to point out that the Random Field Ising Model of interaction between individual firms can lead to the microeconomic hysteresis loops which are not perfectly rectangular \cite{bouchaud2013} and, as shown in  \cite{dahmen1996}, are similar in shape to the hysteresis loop shown in Fig. \ref{fig_stoner}. This brings additional credence to the use of such loops for the description of microeconomic hysteresis.

The microeconomic hysteresis loop represented by Fig. \ref{fig_stoner} can be described by the following formula:
\begin{equation}
\hat{\Gamma}_{\alpha ,\beta} u =
{1 \over 2 }
\Bigl( f_{\alpha ,\beta}^{-} (u) - f_{\alpha ,\beta}^{+} (u)  \Bigr)
\hat{\gamma}_{\alpha ,\beta} u
+
{1 \over 2 }
\Bigl( f_{\alpha ,\beta}^{-} (u) + f_{\alpha ,\beta}^{+} (u)  \Bigr)
\ .
\label{eq2.69}
\end{equation}
Indeed, if input $u$ monotonically varies between $-\infty$ and $\alpha$, then:
\begin{equation}
\hat{\gamma}_{\alpha\beta} u(t) = -1 \ .
\label{eq1.2.new}
\end{equation}
This means that according to formula (\ref{eq2.69}) one obtains:
\begin{equation}
\hat{\Gamma}_{\alpha ,\beta} u =
{1 \over 2 }
\Bigl( f_{\alpha ,\beta}^{+} (u) - f_{\alpha ,\beta}^{-} (u)  \Bigr)
+
{1 \over 2 }
\Bigl( f_{\alpha ,\beta}^{-} (u) + f_{\alpha ,\beta}^{+} (u)  \Bigr)
\ ,
\label{eq2.69.min.a}
\end{equation}
which leads to
\begin{equation}
\hat{\Gamma}_{\alpha ,\beta} u = f_{\alpha ,\beta}^{+} (u)
\ .
\label{eq2.69.min.b}
\end{equation}
Similarly, if input $u$ monotonically varies between $\beta$  and $+\infty$, then:
\begin{equation}
\hat{\gamma}_{\alpha\beta} u(t) = +1 \ .
\label{eq1.1.new}
\end{equation}
This means that according to formula (\ref{eq2.69}) we have:
\begin{equation}
\hat{\Gamma}_{\alpha ,\beta} u =
{1 \over 2 }
\Bigl( f_{\alpha ,\beta}^{-} (u) - f_{\alpha ,\beta}^{+} (u)  \Bigr)
+
{1 \over 2 }
\Bigl( f_{\alpha ,\beta}^{-} (u) + f_{\alpha ,\beta}^{+} (u)  \Bigr)
\ ,
\label{eq2.69.plus.a}
\end{equation}
which leads to
\begin{equation}
\hat{\Gamma}_{\alpha ,\beta} u = f_{\alpha ,\beta}^{-} (u)
\ .
\label{eq2.69.plus.b}
\end{equation}
It is interesting to point out that for the types of hysteresis loops shown in Fig. \ref{fig_stoner} the binary actions (such as, hiring-firing) may occur continuously as represented by the two curves $f_{\alpha ,\beta}^{+} (u)$ and $f_{\alpha ,\beta}^{-} (u)$, while in the case of rectangular loops these actions occur only for two distinct values of input $\alpha$ and $\beta$.

Let us next consider the aggregation of microeconomic hysteresis loops show in Fig. \ref{fig_stoner} to the macroeconomic level. By assuming that there exist many firms whose microeconomic hysteresis are described by formula (\ref{eq2.69}), the above aggregation leads to the following expression for macroeconomic output $f(t)$:
\begin{equation}
f(t) =
\iint\limits_{\alpha \geq \beta}
\hat{\Gamma}_{\alpha, \beta} u(t) d\alpha d\beta
\ .
\label{eq_Gamma1}
\end{equation}
By substituting expression (\ref{eq2.69}) into equation (\ref{eq_Gamma1}), one obtains:
\begin{equation}
f(t) =
\iint\limits_{\alpha \geq \beta}
\mu \bigl( \alpha, \beta , u(t) \bigr)
 \hat{\gamma}_{\alpha, \beta} u(t) d\alpha d\beta
 \ + \
F \bigl( u(t) \bigr)
\ ,
\label{eq_Gamma2}
\end{equation}
where
\begin{equation}
\mu \bigl( \alpha, \beta , u(t) \bigr)
=
{1 \over 2 }
\Bigl( f_{\alpha ,\beta}^{-} \bigl( u(t) \bigr) - f_{\alpha ,\beta}^{+} \bigl( u(t) \bigr)  \Bigr)
\ ,
\label{eq_Gamma3}
\end{equation}
%
%
\begin{equation}
F \bigl( u(t) \bigr)
=
{1 \over 2 }
\iint\limits_{\alpha \geq \beta}
\Bigl(
f_{\alpha ,\beta}^{-} \bigl( u(t) \bigr)
+
f_{\alpha ,\beta}^{+} \bigl( u(t) \bigr)
\Bigr)
d\alpha d\beta
\ .
\label{eq_Gamma5}
\end{equation}
As a result, we have arrived at the generalized Preisach model represented by expressions (\ref{eq_Gamma2})-(\ref{eq_Gamma5}).

\begin{figure}[t]
\centerline{\includegraphics[width=7cm]{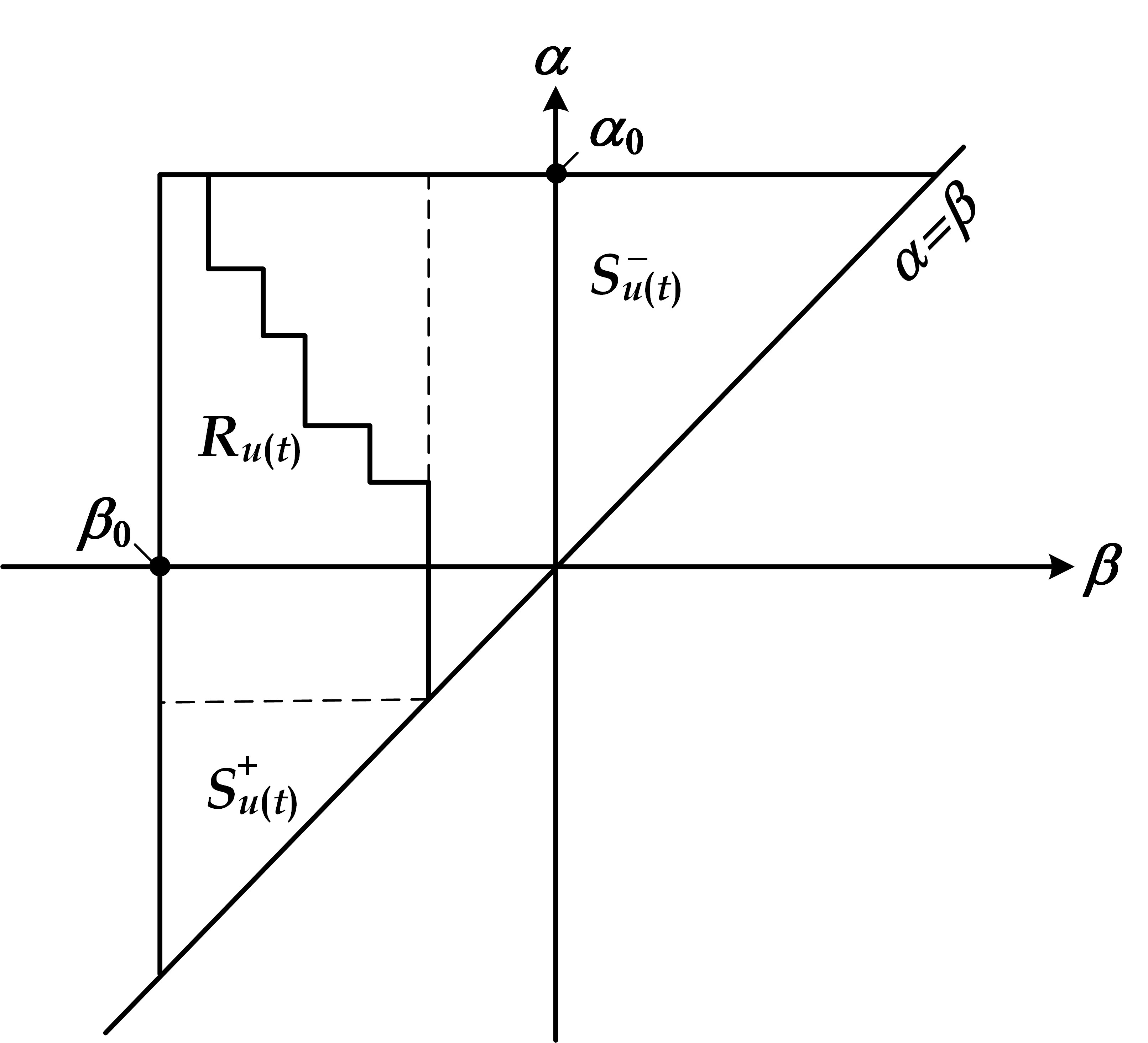}}
\caption{Geometric illustration of the modification of the Preisach model.}
\label{fig_10}
\end{figure}

It is apparent from formula (\ref{eq_Gamma2}) and (\ref{eq_Gamma5}) that the last term in formula (\ref{eq_Gamma2}) is fully reversible, and all irreversibility due to hysteresis is represented by the first term of (\ref{eq_Gamma2}). It turns out that the same first term in expression (\ref{eq_Gamma2}) contains also an additional reversible part. To demonstrate this, we shall write the first term in formula (\ref{eq_Gamma2}) as follows:
\begin{equation}
g(t) =
\iint\limits_{T}
\mu \big( \alpha, \beta , u(t) \big)
\hat{\gamma}_{\alpha, \beta} u(t) d\alpha d\beta
\ ,
\label{eq2.1}
\end{equation}
\noindent
where $T$ is the triangle specified by inequalities $\beta_0 \leq \beta \leq \alpha \leq \alpha_0$, outside of which the weight function $\mu (\alpha, \beta , u(t) )$ is equal to zero. This triangle can be subdivided into three sets $S_{u(t)}^{+}$, $R_{u(t)}$ and $S_{u(t)}^{-}$ (see Fig. \ref{fig_10}), that are defined as follows:
\begin{align}
(\alpha, \beta ) \in S_{u(t)}^{+} ,
&
\ \ \ \ \ {\rm if} \ \ \beta_0 \leq \beta \leq \alpha \leq u(t) ,
\label{eq2.2}
\\
(\alpha, \beta ) \in R_{u(t)} ,
&
 \ \ \ \ \ {\rm if} \ \
\beta_0 \leq \beta \leq u(t) , \ \ u(t) \leq \alpha \leq \alpha_0 ,
\label{eq2.3}
\\
(\alpha, \beta ) \in S_{u(t)}^{-} ,
&
\ \ \ \ \ {\rm if} \ \
u(t) \leq \beta \leq \alpha \leq \alpha_0 .
\label{eq2.4}
\end{align}

\noindent
By using this subdivision of the triangle $T$, it can be shown that $g(t)$ can be expressed as follows:
\begin{equation}
g(t) =
\iint\limits_{R_{u(t)}}
\mu \bigl( \alpha, \beta , u(t) \bigr)
\hat{\gamma}_{\alpha, \beta} u(t) d\alpha d\beta
\ + \
G \bigl( u(t) \bigr)
\ ,
\label{eq2.12}
\end{equation}
\noindent
where, $G(u(t))$ is fully reversible and is given by the following formula:
\begin{equation}
G \bigl( u(t) \bigr) =
\iint\limits_{S_{u(t)}^{+}} \mu (\alpha, \beta , u(t) ) d\alpha d\beta
-
\iint\limits_{S_{u(t)}^{-}} \mu (\alpha, \beta , u(t) ) d\alpha d\beta
\ .
\label{eq2.11}
\end{equation}
\noindent
From (\ref{eq2.1}) and (\ref{eq2.12}) , it can be concluded that expression (\ref{eq_Gamma2}) can be written as:
\begin{equation}
f(t) =
\iint\limits_{R_{u(t)}}
\mu \bigl( \alpha, \beta , u(t) \bigr)
\hat{\gamma}_{\alpha, \beta} u(t) d\alpha d\beta
\ + \
G \bigl( u(t) \bigr)
\ + \
F \bigl( u(t) \bigr)
\ .
\label{Gamma2.mod1}
\end{equation}

It is evident now that the irreversible and reversible components of the hysteresis nonlinearity described by the generalized Preisach model (\ref{eq_Gamma2}) are completely separated. {\bf Since the irreversible part is  most essential for the analysis of economic hysteresis}, then the reversible part can be omitted which results in the formula:
\begin{equation}
\tilde{f} (t) =
\iint\limits_{R_{u(t)}}
\mu \bigl( \alpha, \beta , u(t) \bigr)
\hat{\gamma}_{\alpha, \beta} u(t) d\alpha d\beta
\ ,
\label{Gamma2.mod2}
\end{equation}
\noindent
where $\tilde{f} (t)$ is the irreversible part of $f(t)$.

It is very important to point out that the classical Preisach model (\ref{eq1.3}) also contains both reversible and irreversible components. These components can be separated by using the same line of reasoning as in the derivation of formula (\ref{eq2.12}) from formula (\ref{eq2.1}). In other words, the classical Preisach model can be written as follows:
\begin{equation}
f(t) =
\iint\limits_{R_{u(t)}}
\mu \bigl( \alpha, \beta \bigr)
\hat{\gamma}_{\alpha, \beta} u(t) d\alpha d\beta
\ + \
G \bigl( u(t) \bigr)
\ ,
\label{eq2.12.cpm1}
\end{equation}
\noindent
where
\begin{equation}
G \bigl( u(t) \bigr) =
\iint\limits_{S_{u(t)}^{+}} \mu (\alpha, \beta  ) d\alpha d\beta
-
\iint\limits_{S_{u(t)}^{-}} \mu (\alpha, \beta ) d\alpha d\beta
\ .
\label{eq2.11.cpm}
\end{equation}
Again, if one is interested only in the irreversible component of hysteresis then
\begin{equation}
\tilde{f}(t) =
\iint\limits_{R_{u(t)}}
\mu \bigl( \alpha, \beta \bigr)
\hat{\gamma}_{\alpha, \beta} u(t) d\alpha d\beta
\ .
\label{eq2.12.cpm2}
\end{equation}
\noindent
where $\tilde{f} (t)$ is the irreversible component of $f(t)$ within the context of the classical Preisach model.

Next, we discuss another important issue. A unique feature of economic hysteresis in comparison with hysteresis in natural sciences is due to its ever changing background as a result of the continuous economic evolution. It turns out that the generalized Preisach model of hysteresis can be very useful to account for the continuous evolution of the economy.

It is apparent from formulas (\ref{Gamma2.mod2}) and (\ref{eq2.12.cpm2}) that the main difference between the generalized and the classical Preisach models is the dependence of $\mu$ on $u(t)$ in the case of the generalized model. It turns out that this dependence can be used to account for the evolution of the economy. Indeed, the
switching thresholds $\alpha$ and $\beta $ of microeconomic rectangular hysteresis loops used in the classical Preisach model (\ref{eq1.3}) may {\it  not} remain constant as a result of changes in the economy. These switching values may be shifted as the economy evolves. This may be expressed as follows:
\begin{equation}
\beta^{'} = \beta- g_1 \bigl( u(t) \bigr)       \ , \ \
\alpha^{'} =  \alpha - g_2 \bigl( u(t) \bigr)   \ ,
\label{eq3.45.mod1}
\end{equation}
where $g_2 (u) \leq g_1 (u)$ represent the input dependent shifts in the "up" and "down" threshold values of the rectangular hysteresis loop, respectively. Consequently, the classical Preisach model (\ref{eq2.12.cpm2}) can be written as follows:
\begin{equation}
\tilde{f} (t) =
\iint\limits_{R_{u(t)}}
\mu (\alpha, \beta )
\hat{\gamma}_{\alpha  - g_2 ( u(t) ) , \beta - g_1 ( u(t) ) } u(t) d\alpha d\beta
.
\label{eq2.1.shift1}
\end{equation}
We next consider formula (\ref{eq3.45.mod1}) as a change of variables from $\alpha$ and $\beta$ to $\alpha^{'}$ and $\beta^{'}$, respectively. This leads to the modification of the last formula:
\begin{equation}
\tilde{f} (t) =
\iint\limits_{R_{u(t)}^{'}}
\mu \bigl( \alpha^{'} + g_2 \bigl( u(t) \bigr)  , \beta^{'} + g_1 \bigl( u(t) \bigr)  \bigr)
\hat{\gamma}_{\alpha^{'} , \beta^{'} } u(t) d\alpha^{'} d\beta^{'}
.
\label{eq2.1.shift2}
\end{equation}
Now, it is clear that the weight function $\mu$ is dependent on the input $u(t)$. By introducing the new notation for the weight function,
\begin{equation}
\tilde{\mu} \bigl( \alpha^{'}  , \beta^{'} , u(t) \bigr)
\equiv
\mu \bigl( \alpha^{'} + g_2 \bigl( u(t) \bigr)  , \beta^{'} + g_1 \bigl( u(t) \bigr)  \bigr)
,
\label{eq2.1.shift3}
\end{equation}
formula (\ref{eq2.1.shift2}) can be written as follows:
\begin{equation}
\tilde{f} (t) =
\iint\limits_{R_{u(t)}^{'}}
\tilde{\mu} \bigl( \alpha^{'}  , \beta^{'} , u(t)  \bigr)
\hat{\gamma}_{\alpha^{'} , \beta^{'} } u(t) d\alpha^{'} d\beta^{'}
,
\label{eq2.1.shift4}
\end{equation}
which is mathematically identical to the expression of the generalized Preisach model. This clearly demonstrates that when the evolution of the economy results in shifts in the thresholds of binary actions, the resultant economic hysteresis may be represented by the generalized Preisach model where the weight function depends on the input $u(t)$.
A similar reasoning may be used to account for the shift of the thresholds $\alpha$ and $\beta$ due to the evolution of the economy in the context of the generalized model.

\begin{figure}[h]
\centerline{\includegraphics[width=8.0cm]{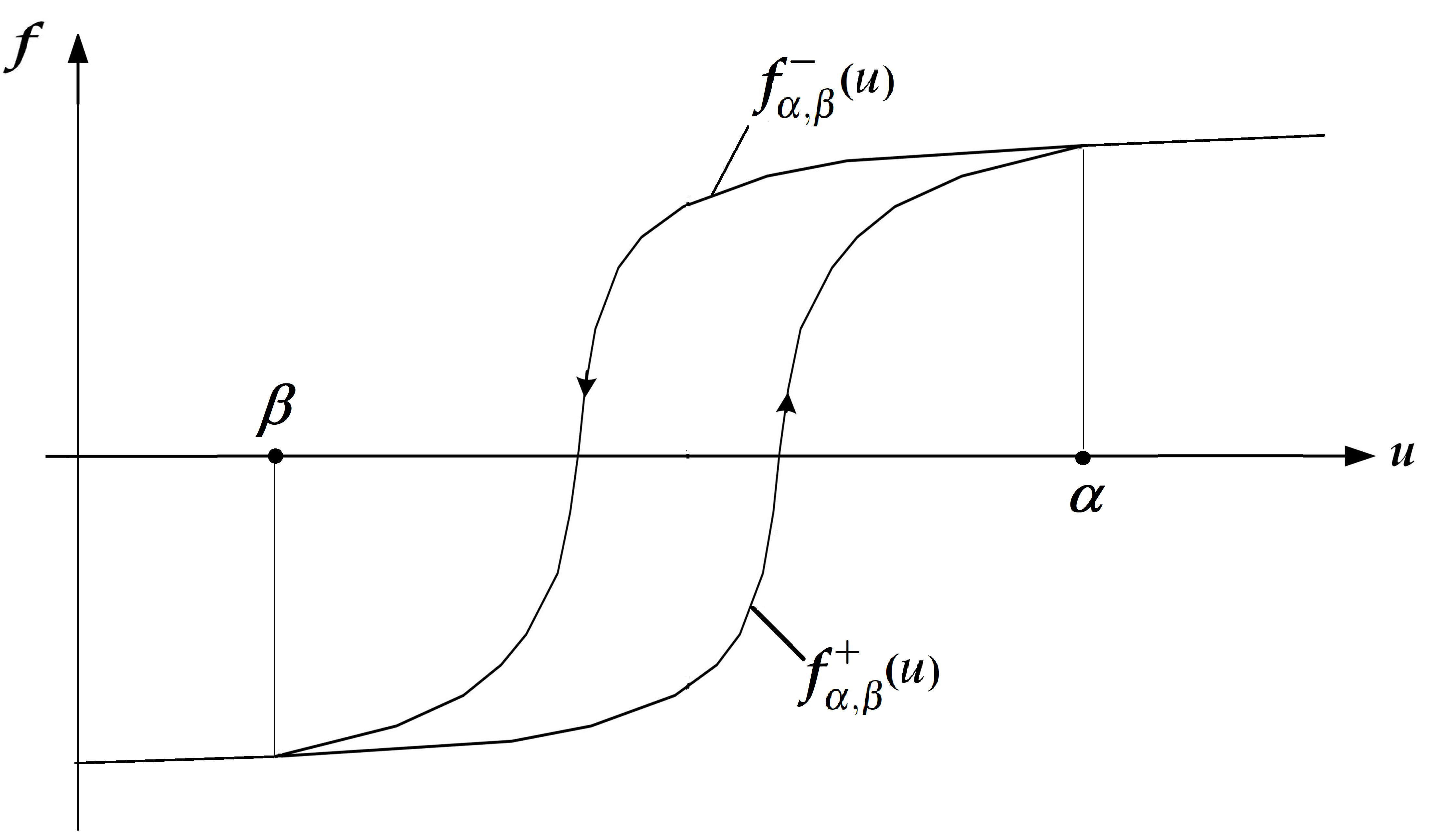}}
\caption{A general microeconomic hysteresis loop.}
\label{fig_realistic}
\end{figure}

To conclude this section, it is interesting to point out that the generalized Preisach model can be also viewed as a macroeconomic aggregation of microeconomic hysteresis represented by Fig. \ref{fig_realistic}. This is the case because formula (\ref{eq2.69}) is quite general and it is valid for the hysteresis loop shown in Fig. \ref{fig_realistic}. For this hysteresis loop $\alpha$ and $\beta$ are the values of input at which ascending and descending branches merge together, while functions $f^+$ and $f^-$ correspond to the above branches, respectively. The justification of formula (\ref{eq2.69}) for the hysteresis loop shown in Fig. \ref{fig_realistic} as well as the derivation of the generalized Preisach model for this case literally repeats our reasoning presented above for the case of hysteresis loop shown in Fig. \ref{fig_stoner}.

\section{\label{sec:level1}Properties of the Generalized Preisach Model}

\begin{figure}[t]
\centerline{\includegraphics[width=7.5cm]{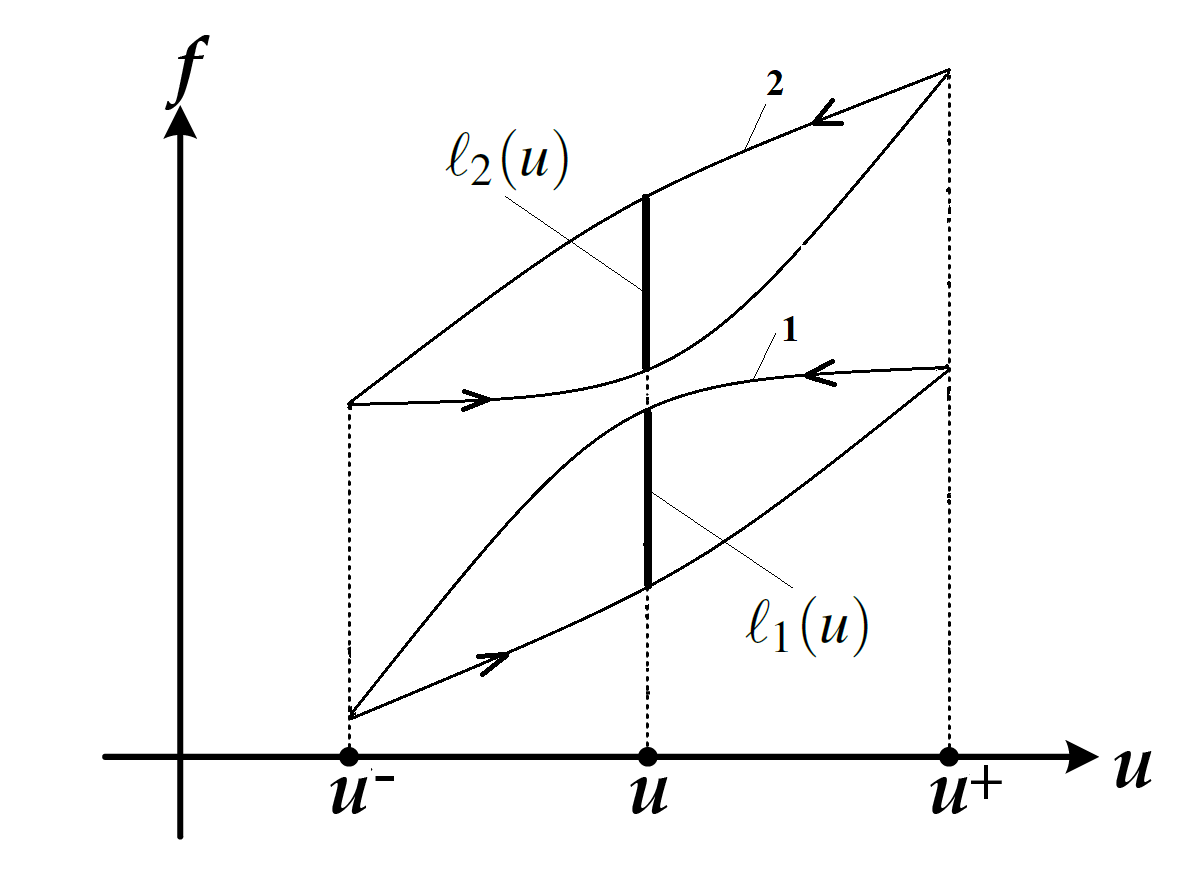}}
\caption{Geometric illustration of the Property of Equal Vertical Chords.}
\label{fig_12}
\end{figure}

It can be shown that the same Erasure Property is valid for the generalized model as it is for the Classical Preisach model (for instance, see \cite{neural2020}). This is because the formation of the staircase interface $L(t)$ and its modifications by input variations are still controlled by the switching of rectangular loops $\hat{\gamma}_{\alpha, \beta}$. This leads to the same two rules of formation of $L(t)$ as in the case of the classical Preisach model (\ref{eq1.3}). On the other hand, the Congruency Property of the classical Preisach model needs to be modified in the case of the generalized model. Indeed, consider two inputs $u_1(t)$ and $u_2 (t)$ that may have different past histories. However, starting from some instant of time these inputs vary back and forth between the same two consecutive values $u^-$ and $u^+$. As a result, two distinct hysteresis loop will be formed as shown in Fig. \ref{fig_12}. Then, the following property is valid for such hysteresis loops:

\vfill
\medskip\noindent
{\bf PROPERTY OF EQUAL VERTICAL CHORDS} \ {\it Hysteresis loops resulting from back-and-forth input variations between the same two consecutive input extrema have equal vertical chords (output increments) for the same input values but they are not congruent} (see Fig. \ref{fig_12}).

\begin{figure}[h]
\centerline{\includegraphics[width=7cm]{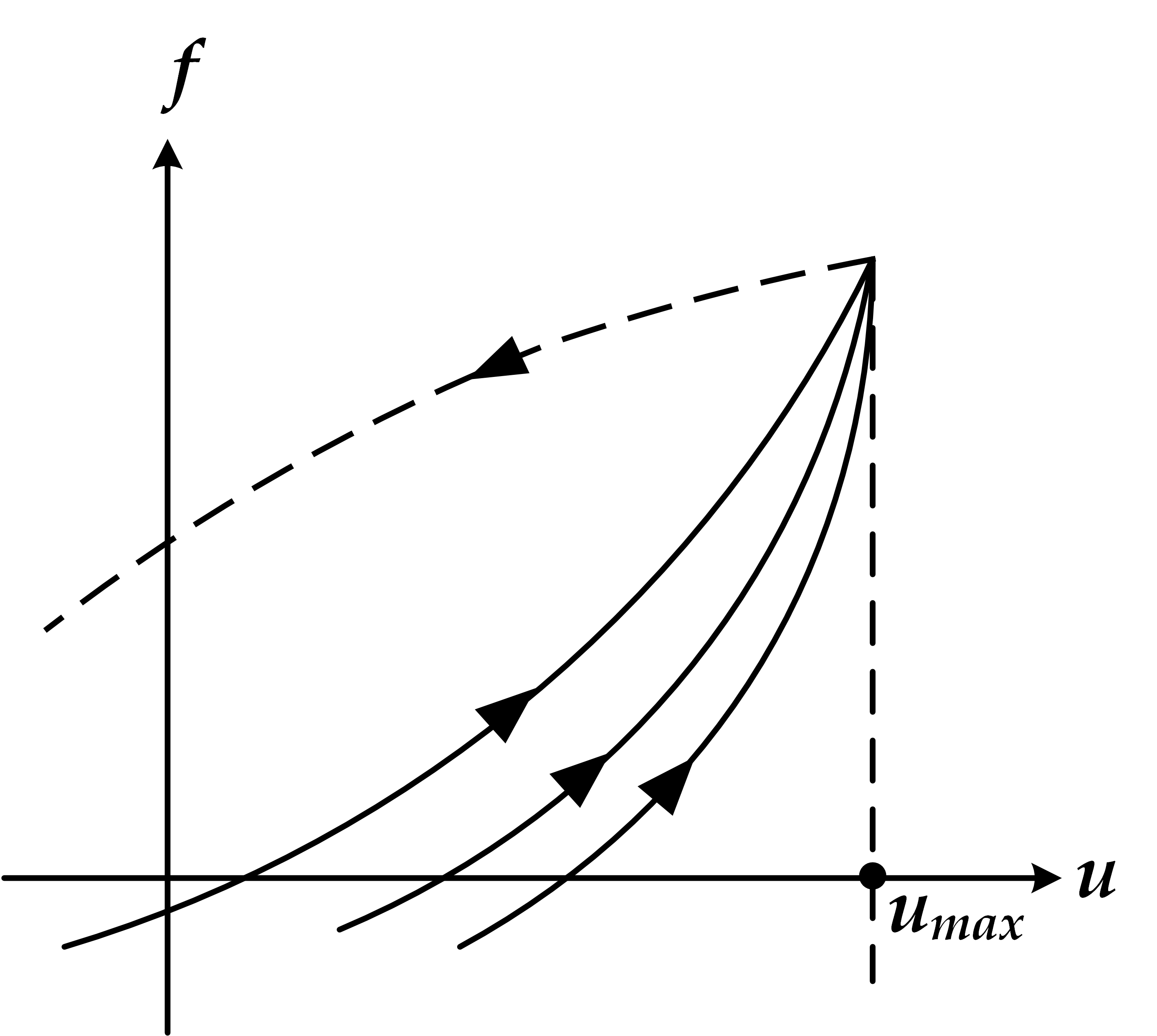}}
\caption{Illustration of branching of the classical Preisach model.}
\label{fig_13}
\end{figure}

\begin{figure}[h]
\centerline{\includegraphics[width=7cm]{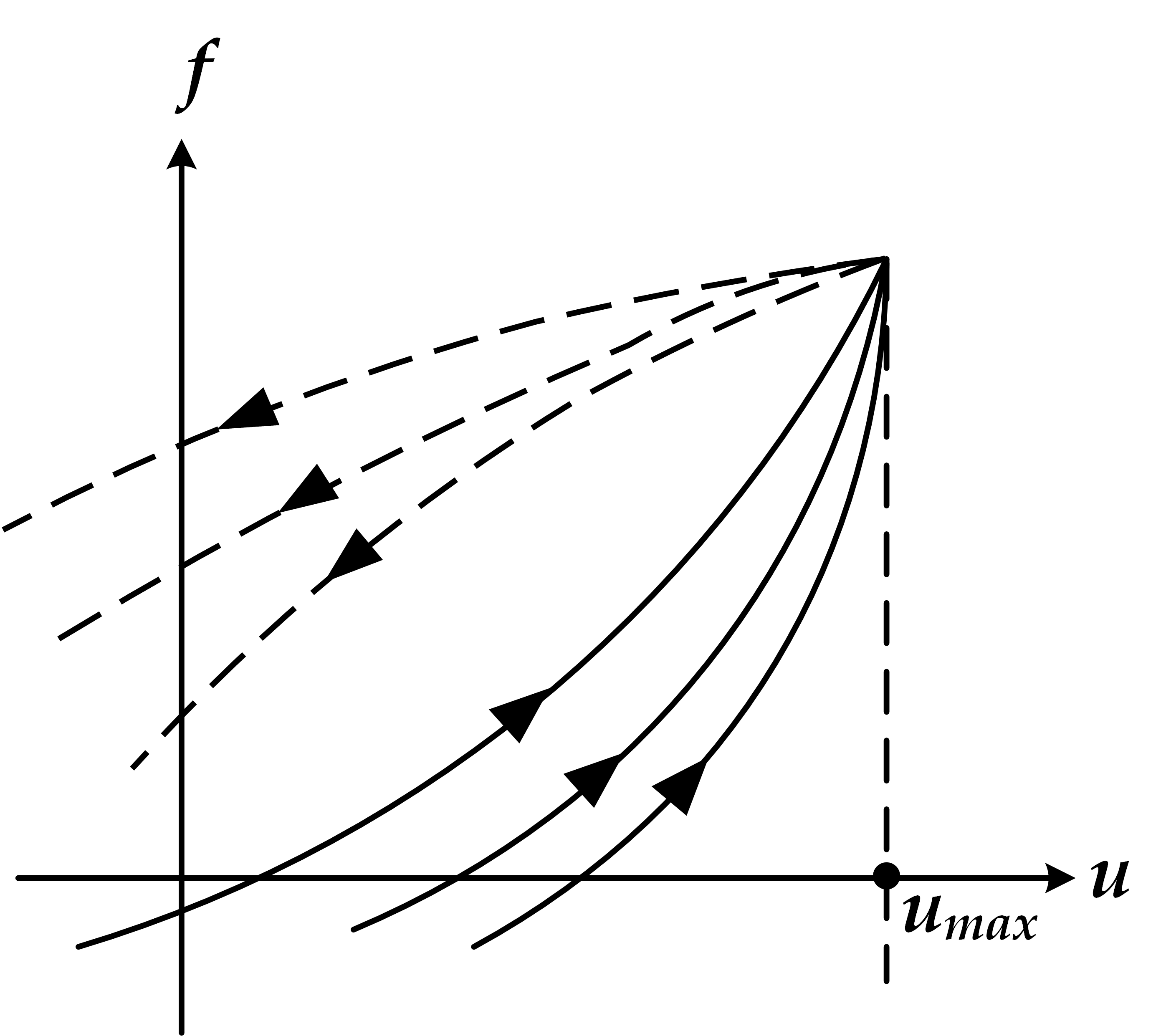}}
\caption{Illustration of branching of the generalized Preisach model.}
\label{fig_14}
\end{figure}

\begin{figure}[t]
\centerline{\includegraphics[width=7.5cm]{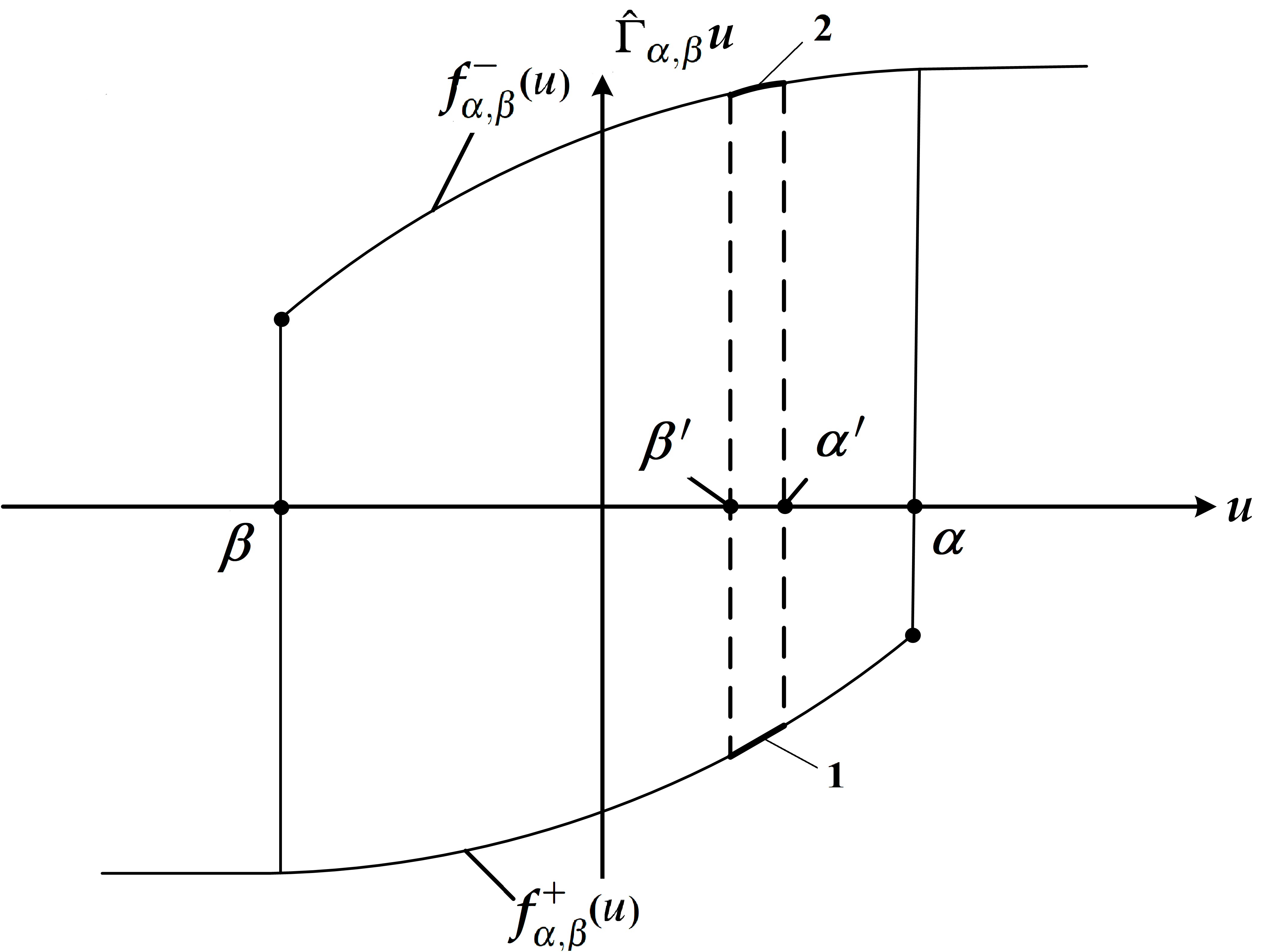}}
\caption{Illustration of incongruent (minor) loops with zero length vertical chords for microeconomic hysteresis.}
\label{fig_stoner_zero}
\end{figure}

\begin{figure}[h]
\centerline{\includegraphics[width=8cm]{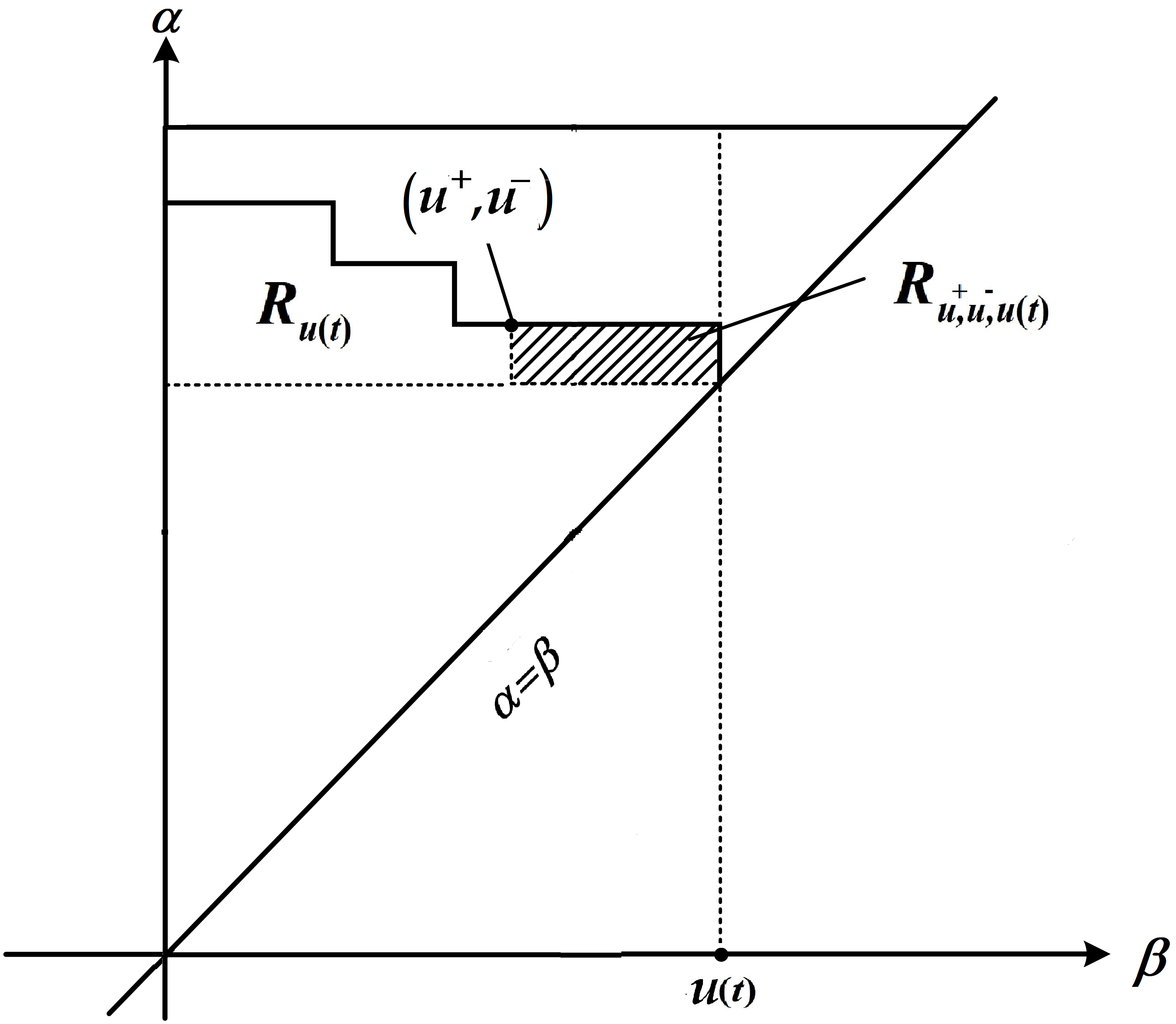}}
\caption{Illustration of rectangles  $R_{u^{+},u^{-},u(t)}$ and $R_{u(t)}$ for the generalized model.}
\label{fig_rabu}
\end{figure}

\medskip\noindent
Namely, the equality $\ell_1 (u) = \ell_2 (u)$ holds according to the above property.

It turns out that the above two properties are very important. This is because they are the very foundation of the following representation theorem.

\medskip\noindent
{\bf REPRESENTATION THEOREM} {\it The erasure property and the property of equal vertical chords for hysteresis loops corresponding to the same input extrema constitute the necessary and sufficient conditions for the representation of a hysteresis nonlinearity by the generalized Preisach model}.

It is clear that congruent loops have equal vertical chords, while hysteresis loops with equal vertical chords are not necessarily congruent. This suggests that the stated representation theorem is more general than the corresponding representation theorem for the classical Preisach model. This further implies that the generalized Preisach model has predictive powers within the wider scope of applicability as compared to the classical Preisach model.

It is stressed in Section \ref{sec:whatishysteresis}, that branching is the essence of hysteresis. For this reason, it is interesting to compare the branching properties of the classical and generalized Preisach models of hystersis. In this respect, it is important to make the following observation in order to appreciate the extent to which the generalized Preisach model is more advanced than the classical model. The classical Preisach model represents hysteresis nonlinearities that, at any {\it reversal} point, have (regardless of past history) only one branch starting from this point (see Fig. \ref{fig_13}). This can be shown by employing the congruency property of the classical Preisach model. On the other hand, the generalized Preisach model describes hysteresis nonlinearities with the property that at any reversal point there are {\it infinite} possible branches starting at this point (see Fig. \ref{fig_14}). A particular realization of these branches is determined by a particular past history. This observation shows that the generalized Preisach model is endowed with a more sophisticated mechanism of branching as compared to the classical Preisach model. For this reason, by extending the terminology of \cite{amable1991} introduced earlier, one may say that the generalized Preisach model describes "super-strong" hysteresis.

To further demonstrate the broader applicability of the generalized Preisach model in comparison with the classical Preisach model, we shall return to the discussion of the hysteresis loop shown in Fig. \ref{fig_stoner_zero}. It turns out that this loop cannot be represented by the classical Preisach model. The reason is that degenerate loops $1$ and $2$ traced as the input varies back and forth between $\alpha^{'}$ and $\beta^{'}$ are not congruent. Thus, the "weak" hysteresis of the loop shown in Fig. \ref{fig_stoner_zero} cannot be represented by the aggregation of rectangular loops which are the building blocks of the "strong" hysteresis of the classical Preisach model. On the other hand, the hysteresis loop shown in Fig. \ref{fig_stoner_zero} can be described by the generalized Preisach model. We have already discussed this matter before (see Formula (\ref{eq2.69})). This is also clear from the fact that the property of equal vertical chords is satisfied because all minor loops with $\beta \leq \beta^{'}$ and $\alpha^{'} \leq \alpha$ are degenerate and, consequently, they have equal (zero) vertical chords.

We conclude this section by pointing out that similar to the case of the classical Preisach model, the Property of Equal Vertical Chords may have important economic implications. As previously discussed, in the case of cyclical unemployment hysteresis, the length $\ell$ of vertical chords of hysteresis loops can be viewed as a measure of sluggishness of economic recovery. This sluggishness is {\it intrinsic} and it is a consequence of hysteresis branching. It can be shown that the length $\ell$ of vertical chords is given by the formula:
\begin{equation}
\ell =
2 \iint\limits_{R_{u^{+},u^{-},u(t)}}
\mu \big( \alpha , \beta , u(t) \big) d\alpha d\beta
\ ,
\label{eq2.50_rev1}
\end{equation}
where $R_{u^{+},u^{-},u(t)}$ is the rectangle shown in Fig. \ref{fig_rabu}. As before, this integral can be evaluated in terms of the microeconomic employment data of binary agents whose triggering values $\alpha$ and $\beta$ are such that the points $(\alpha ,\beta )$ are confined within the rectangle $R_{u^{+},u^{-},u(t)}$.

According to the Property of Equal Vertical Chords, the vertical chords are the same for all hysteresis loops corresponding to the same input extrema $u^{-}$ and $u^{+}$. In this respect, {\bf the sluggishness of the economic recovery is universal} and does not depend on the history of input variations prior to the commencement of the economic cycle determined by inputs $u^-$ and $u^+$. According to the generalized model, it is remarkable that this universality is not affected by the economic evolution prior to the economic cycle. Furthermore, this universality exists despite the fact that the output values corresponding to $u^{-}$ and $u^{+}$ are history-dependent.

\section{\label{sec:level1}Conclusion}

The paper explores the mathematical modeling of macroeconomic hysteresis by using Preisach models. The main contributions of the paper can be summarized as follows.

A general definition of hysteresis as history-dependent branching is presented and the relevance of this definition to macroeconomic hysteresis is stressed. It is pointed out that the origin of history-dependent branching in macroeconomics is due to the aggregation of binary microeconomics hysteresis (such as hiring-firing, buying-selling, etc.) occurring at various values of economic input. Such binary actions can be modeled by rectangular hysteresis loops. This reveals that the origin of macroeconomic hysteresis is intimately related to the mathematical structure of the classical Preisach model. This model is presented in the paper and its basic properties related to economic hysteresis are discussed.

Subsequently, the generalized Preisach model of hysteresis is introduced. This model is constructed as a macroeconomic aggregation of a more realistic microeconomic hysteresis than in the case of the classical Preisach model. It is demonstrated that this model is endowed with a more general and sophisticated mechanism of branching as compared to the classical Preisach model. Furthermore, it is shown that the generalized model may account for the continuous evolution of the economy and its effect on hysteresis.

It is demonstrated in the paper that sluggish recoveries are due to hysteresis branching. Indeed, the lengths of the vertical  chords of hysteresis loops corresponding to cyclical unemployment hysteresis can be viewed as measures of economic recovery sluggishness. These lengths can be easily computed by using the formalism of the Preisach models. It is remarkable that the above lengths of vertical chords do not depend on the history of input variations. In this sense, these vertical chords provide universal measures of economic recovery sluggishness. These facts clearly suggest that the analysis of economic hysteresis as branching may provide more useful information than the traditional analysis based on the conventional time-series method.

\section*{Acknowledgement}

The authors would like to acknowledge Professor Rod Cross for reaching to them and enhancing their interest in the mathematical modeling of economic hysteresis.

\bibliographystyle{unsrt}
\bibliography{econ_references}

\end{document}